\documentclass[sigconf,screen]{acmart}

\usepackage{hyperref}
\usepackage{todonotes}
\usepackage{enumitem}
\usepackage{hyphenat}
\usepackage{calc}
\usepackage{natbib}
\usepackage{amsmath}
\usepackage{graphicx}
\usepackage{mathtools}
\usepackage{booktabs}
\usepackage{url}
\usepackage{makecell}
\usepackage{listings}
\usepackage{acmart-taps}

\definecolor{kwblue}{rgb}{0, 0.113, 0.64}
\definecolor{commentgray}{rgb}{0.47, 0.47, 0.47}
\definecolor{stringgreen}{rgb}{0.066, 0.425, 0.07}
\definecolor{dkgreen}{rgb}{0,0.6,0}
\definecolor{gray}{rgb}{0.5,0.5,0.5}
\definecolor{mauve}{rgb}{0.58,0,0.82}

\lstdefinelanguage{JavaScript}{
  keywords={break, case, catch, continue, debugger, default, delete, do, else, false, finally, for, function, if, in, instanceof, new, null, return, switch, this, throw, true, try, typeof, var, void, while, with},
  morecomment=[l]{//},
  morecomment=[s]{/*}{*/},
  morestring=[b]',
  morestring=[b]",
  ndkeywords={class, export, boolean, throw, implements, import, this},
  keywordstyle=\color{kwblue}\bfseries,
  ndkeywordstyle=\color{darkgray}\bfseries,
  identifierstyle=\color{black},
  commentstyle=\color{commentgray}\ttfamily,
  stringstyle=\color{stringgreen}\ttfamily,
  sensitive=true
}

\lstdefinelanguage{Prompt}{
  keywords={class, static, user, agent, parsed},
  morecomment=[s]{<}{>},
  keywordstyle=\color{kwblue}\bfseries,
  ndkeywordstyle=\color{darkgray}\bfseries,
  identifierstyle=\color{black},
  commentstyle=\color{commentgray}\ttfamily,
  stringstyle=\color{stringgreen}\ttfamily,
  sensitive=true
}

\aptLtoX[graphic=no,type=html]{}{\renewcommand{\texttt}[1]{%
  \begingroup
  \ttfamily
  \hyphenchar\font=-1\relax
  \expandafter\texttthelper#1\relax
  \endgroup
}}

\makeatletter
\aptLtoX[graphic=no,type=html]{}{\newcommand{\texttthelper}[1]{%
    \ifx\relax#1\else
    \ifx(#1\allowbreak\fi
    \ifx[#1\allowbreak\fi
    \ifx]#1\allowbreak\fi
    \ifx:#1\allowbreak\fi
    #1%
    \expandafter\texttthelper
  \fi
}}
\makeatother

\copyrightyear{2024}
\acmYear{2024}
\setcopyright{rightsretained}
\acmConference[CHI '24]{Proceedings of the CHI Conference on Human Factors in Computing Systems}{May 11--16, 2024}{Honolulu, HI, USA}
\acmBooktitle{Proceedings of the CHI Conference on Human Factors in Computing Systems (CHI '24), May 11--16, 2024, Honolulu, HI, USA}
\acmDOI{10.1145/3613904.3642517}
\acmISBN{979-8-4007-0330-0/24/05}

\newcommand{\mr}{ReactGenie}
\newcommand{\Mrd}{Natural Language Programming Language}
\newcommand{\mrd}{NLPL}

\begin{document}
\sloppy
\title[\mr{}]{\mr{}: A Development Framework for Rich Multimodal Interactions Using Large Language Models}
\author{Jackie Junrui Yang} 
\orcid{0000-0002-2064-5231}
\email{jackiey@stanford.edu}
\affiliation{%
    \institution{Stanford University}
    \city{Stanford}
    \state{CA}
    \country{USA}
}
\author{Yingtian Shi}
\orcid{0000-0001-8733-7041}
\affiliation{%
    \institution{Tsinghua University}
    \city{Beijing}
\country{China}
}
\author{Yuhan Zhang}
\orcid{0009-0000-7720-9329}
\email{zhangyh@stanford.edu}
\affiliation{%
    \institution{Stanford University}
    \city{Stanford}
    \state{CA}
    \country{USA}
}
\author{Karina Li}
\orcid{0009-0002-2711-5185}
\email{karinali@stanford.edu}
\affiliation{%
    \institution{Stanford University}
    \city{Stanford}
    \state{CA}
    \country{USA}
}
\author{Daniel Wan Rosli}
\orcid{0009-0005-5625-8312}
\email{danwr@stanford.edu}
\affiliation{%
    \institution{Stanford University}
    \city{Stanford}
    \state{CA}
    \country{USA}
}
\author{Anisha Jain}
\orcid{0009-0003-4819-529X}
\email{anishaj037@gmail.com}
\affiliation{%
    \institution{Independent Researcher}
    \country{USA}
}
\author{Shuning Zhang}
\orcid{0000-0002-4145-117X}
\email{zhang-sn19@mails.tsinghua.edu.cn}
\affiliation{%
    \institution{Tsinghua University}
    \city{Beijing}
    \country{China}
}
\author{Tianshi Li}
\email{tia.li@northeastern.edu}
\orcid{0000-0003-0877-5727}
\affiliation{%
    \institution{Northeastern University}
    \city{Boston}
    \state{MA}
    \country{USA}
}
\author{James A. Landay}
\email{landay@stanford.edu}
\orcid{0000-0003-1520-8894}

\affiliation{%
    \institution{Stanford University}
    \streetaddress{353 Jane Stanford Way}
    \city{Stanford}
    \state{CA}
    \country{USA}
    \postcode{94305–2004}
}
\author{Monica S. Lam}
\email{lam@cs.stanford.edu}
\orcid{0000-0002-7626-6468}
\affiliation{%
    \institution{Stanford University}
    \streetaddress{353 Jane Stanford Way}
    \city{Stanford}
    \state{CA}
    \country{USA}
    \postcode{94305–2004}
}

\renewcommand{\shortauthors}{Yang et al.}

\begin{abstract}

By combining voice and touch interactions, multimodal interfaces can surpass the efficiency of either modality alone.
Traditional multimodal frameworks require laborious developer work to support rich multimodal commands where the user's multimodal command involves possibly exponential combinations of actions/function invocations.
This paper presents ReactGenie, a programming framework that better separates multimodal input from the computational model to enable developers to create efficient and capable multimodal interfaces with ease.
ReactGenie translates multimodal user commands into \mrd{} (\Mrd{}), a programming language we created, using a neural semantic parser based on large-language models.
The ReactGenie runtime interprets the parsed \mrd{} and composes primitives in the computational model to implement complex user commands.
As a result, ReactGenie allows easy implementation and unprecedented richness in commands for end-users of multimodal apps.
Our evaluation showed that 12 developers can learn and build a non-trivial ReactGenie application in under 2.5 hours on average.
In addition, compared with a traditional GUI, end-users can complete tasks faster and with less task load using ReactGenie apps.
\end{abstract}





\maketitle

\section{Introduction}

\begin{figure*}
    \centering
    \vspace{-0.5cm}
    \includegraphics[width=\linewidth]{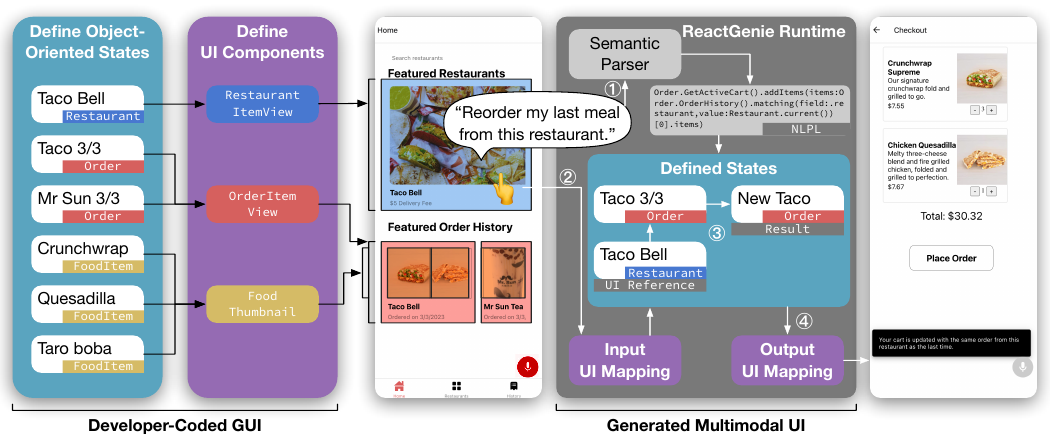}
    \vspace{-0.5cm}
    \caption{
        \mr{} allows developers to easily build multimodal applications by better-separating interfaces (UI components) from computational models (object-oriented state). \enspace
        The demo (two screenshots) shows the user performing a multimodal (speech + touch) command (left screenshot),  with the system executing the command by parsing the voice, understanding the reference in touch, and presenting the user with the appropriate UI interface and text feedback (right screenshot). \enspace
        (Left) \mr{} provides this new yet familiar interface to create a GUI by defining states (data and logic) and UI components (transformation from data to UI representation). \enspace
        (Right) \mr{} automatically generates a natural semantic parser from developer-defined states and generates input and output UI mappings from developer-defined UI components.
        \mr{} can then execute rich multimodal commands by composing the methods and properties of states and presenting the results using existing UI components.
    }
    \Description{
        This figure shows the \mr{} framework:
        The developer programs a GUI through object-oriented states and declarative UI components.
        \mr{} then generates a system that allows the user to interact with the GUI multimodally.
    }
    \label{fig:teaser}
\end{figure*}

Multimodal interactions, combining multiple different input and output modalities, such as touch, voice, and graphical user interfaces (GUIs), offer increased flexibility, efficiency, and adaptability for diverse users and tasks~\cite{10.1016/j.patrec.2013.07.003}.
However, the development of multimodal applications remains challenging for developers due to the complexity of managing multimodal commands and handling the low-level control logic for interactions.
Existing frameworks~\cite{10.1145/1358628.1358881,10.1145/1570433.1570476,10.1145/1979742.1979703,10.1080/088395199117504,10.1145/2070481.2070500,10.48550/arxiv.2103.15975,10.1109/vrw55335.2022.00018} often require developers to manually handle these complexities, significantly increasing development costs and time.
The voice modality, in particular, presents a unique challenge due to the compositionality and expressiveness of natural language.
Sub-par implementations often greatly reduce the expressiveness of these multimodal interfaces.
Various systems~\cite{garg2023,10.48550/arxiv.2209.08655} can automatically handle voice commands by converting them to UI actions, but they are prone to error and do not allow developers to fully control the app's behavior.

The research described in this paper aims to provide developers with a simple programming abstraction (see \autoref{fig:teaser}) by hiding the complexity of natural language understanding and supporting the composition of different modalities automatically.
Our goal is to enable users to access off-screen content/actions and complete tasks that normally involve multiple GUI taps in a single multimodal command, as illustrated in \autoref{fig:interaction}.
This flexibility is achieved with little additional effort from developers compared to traditional GUI apps.
This approach encourages the adoption of multimodal interactions and makes multimodal interactions more accessible to end-users.

This paper presents \mr{}\footnotemark, a declarative programming framework for developing multimodal applications.
\footnotetext{project website (including source code): \url{https://jya.ng/reactgenie}, \url{https://hci.stanford.edu/research/reactgenie/}}
The core concept behind \mr{} is a better abstraction that separates the multimodal input and output interfaces from the underlying computation models.
\mr{} uses an object-oriented state abstraction to represent the computation model of the app and uses declarative UI components to represent the UI\@.
Users' compound multimodal commands are translated into a composition of multiple function calls using large language models (LLMs), e.g., to find the referred object/objects and make the right state change.

Existing declarative UI state management frameworks, such as Redux~\cite{redux}, use a single global state store to manage all of the state changes of the UI\@.
The straightforward way to implement rich multimodal user commands in these existing frameworks is by making many imperative-style function calls.
However, these function calls require the error-prone creation of many intermediate variables to store return values that are then used in the next function call as the programmer traverses the complex state stored in the monolithic object.
These intermediate variables commonly cause missing references to variables when the neural semantic parser  translates the user's natural language input into code~\cite{10.48550/arxiv.2203.12751}.
In contrast, the object-oriented state abstraction in \mr{} encourages componentized classes instead of a single global state store.
The componentized classes result in smaller objects, each equipped with methods for relevant operations.
This design supports multiple chained method calls/property accesses (method chaining) and provides a straightforward representation of the user's command with no need for intermediate variables (as shown in the example \mrd{} command in \autoref{fig:teaser}).
This allows \mr{} to accurately compose the methods and properties of existing states needed for executing rich multimodal commands.

With \mr{}, developers build graphical interfaces using a development workflow similar to a typical React + Redux~\cite{rr} application. 
To add multimodality, the developer simply adds a few annotations to their code and example parses (pairs of expected end-user voice command examples and the corresponding function calls).
These command examples indicate what methods/properties can be used in voice and how.
By using the extracted class definitions and example parses from the developer's state code, 
\mr{} creates a parser that leverages an LLM~\cite{10.48550/arxiv.2005.14165} to translate the user's natural language to a new domain-specific language (DSL), \textit{\mrd{}}, \Mrd{}.
Combined with a custom-designed interpreter, \mr{} can seamlessly handle multimodal commands and present the results in the graphical UI that the developer builds as usual.

As shown in \autoref{fig:teaser} left, developers can define both object-oriented state abstraction classes to handle data changes and UI components that explicitly map the state to the UI\@.
Similar to React, when the user interacts with the app, the app's state will be updated, and the UI will be re-rendered.
What sets \mr{} apart is its unique ability to support rich multimodal input, as shown in \autoref{fig:teaser} right.

The main contributions of this research are as follows:
\begin{itemize}
    \item \mr{}, a multimodal app development framework based on an object-oriented state abstraction that is easy for developers to learn and use and generates apps that support rich multimodal interactions.
    \item A programming language, \mrd{}, used to represent user's multimodal commands.
    This involves the design of the high-level annotation of user-accessible functions, the automatic generation of a natural semantic parser using LLMs that targets \mrd{}, a new DSL for rich multimodal commands, and an interpreter that executes \mrd{}.
    These systems support automatic and accurate handling of natural language understanding in \mr{}.
    \item Evaluations of \mr{}:
    \begin{itemize}
        \item For developers, we demonstrated its expressiveness through building three representative demo apps in different domains, its low development cost by comparing it with GPT-3 function calling, and its usability and learnability through a study with 12 developers successfully building a demo app.
        \item For end-users, we measured the parser accuracy to be 90\% with elicited commands from 50 participants and evaluated the usability of apps built using ReactGenie in a user study with 16 participants. We found users had a reduced cognitive load when using an app with \mr{}-supported multimodal interactions compared to using a graphical user interface (GUI) app. They also preferred the multimodal app to the GUI-based app.
    \end{itemize}
\end{itemize}

\subsection{Targeted Interactions}

\mr{} supports \textit{rich} interactions that are \textit{complex} for current computer systems, but are \textit{intuitive} for users.
One example of a rich multimodal interaction is shown in the center of \autoref{fig:teaser}: the user says, ``Reorder my last meal from this restaurant'' while touching the restaurant displayed on the screen.
Such commands are common in human-to-human communication.
Still, they involve multiple steps (retrieving the history of orders from the restaurant, creating an order, and adding food to the order) for the app. 
These commands are {\em complex} to implement today as they require combining inputs from both modalities and/or composition of different features.

\mr{} supports a typical family of gesture + speech multimodal interactions.
This aligns with one of the categories of speech and gesture multimodal applications proposed by Sharon Oviatt's seminal work~\cite{10.1201/9781420088861}:
The \textit{recognition modes} \mr{} supports are \textit{simultaneous and individual modes}, meaning that \mr{} supports users to use either speech-only interactions, gesture-only interactions, or both interactions at the same time (``What is the last time I ordered from this \textit{[touch on a restaurant]} restaurant'').
The \textit{supported gesture input type} is \textit{touch/pen input}, and the \textit{size of the gesture vocabulary} is a \textit{deictic selection}.
This means that \mr{} focuses on scenarios where the user's gesture input resolves object references through pointing in a multimodal command.
The \textit{size of speech vocabulary} is \textit{arbitrary human sentences}, and the type of linguistic processing is \textit{large-language model processing}.
The last two terms are new types we invented to better describe \mr{}'s support for rich commands and the use of highly generalizable large-language models.
Following Oviatt's original classification, \mr{} would be classified as \textit{large vocabulary} and \textit{statistical language processing}.
\mr{} uses \textit{late semantic fusion} to fuse input from different modalities, 
which means the system integrates and interprets the meaning of inputs from multiple modalities only after each input has been independently processed and understood.

With \mr{}, the developer simply provides a small amount of additional information associated with each input method and function. 
Our system supports the full compositionality of input modalities and functions by automatically translating a user command into one of exponentially many possible action sequences. The richness of user interaction afforded by our system is unprecedented, as traditional multimodal programming frameworks require developers to hard-code every combination of features supported.

\mr{} lets the programmer simply \textit{describe} the functionality of their code, including actions they support and the relationship between UI and data.
This allows \mr{} to handle these rich multimodal commands in arbitrary combinations of actions without requiring direct developer input.
The example in \autoref{fig:teaser} is supported by:
\begin{enumerate}
\item
\mr{} first translates the user's voice command to the \mrd{} code.
For example, the user refers to an element in the UI by voice (``this restaurant''), and the semantic parser generates a special reference \texttt{Restaurant.current()}.
\item
\mr{} extracts the tap point from the UI and uses the UI component code to map the tap point back to a state object \texttt{Restaurant(name: "Taco Bell")}.
\item
With the parsed DSL and UI context, \mr{}'s interpreter can execute the generated \mrd{} using developer-defined states.
It first retrieves the most recent order from ``Taco Bell'', designated as ``Taco 3/3''.
Then, it creates a new order, designated as ``New Taco''.
Finally, the interpreter adds all the food items from ``Taco 3/3'' to ``New Taco'' and returns the new order.
\item
\mr{} passes the return value of the \mrd{} statement to the output UI mapping module.
Because the return value is an \texttt{Order} object, \mr{} searches in the developer's UI component code to find a corresponding representation (Output UI Mapping) to present the result to the user.
\mr{} also generates a text response using the LLM based on the user's input, parsed \mrd{}, and the return value: ``Your cart is updated with the same order from this restaurant as the last time.''
\end{enumerate}

During this process, the \mr{} framework uses its knowledge about the developer's app to automatically understand a multimodal compositional command, compose actions to execute, and find the appropriate interface to present the results to the user.
This pipeline allows \mr{} to handle more commands than prior frameworks with little developer input.

\section{Related work}

In this section, we review related work on multimodal interaction systems, Graphical and Voice UI frameworks, and multimodal interaction frameworks.

\subsection{Multimodal Interaction Systems}

Many researchers have proposed multimodal interaction systems.
The earliest multimodal interaction systems, such as Bolt's ``Put-that-there'', were developed in the 1980s~\cite{10.1145/800250.807503}.
They demonstrated that users can interact with a computer using voice and gestures.
QuickSet~\cite{10.1145/266180.266328} further demonstrated use cases of multimodal interaction on a mobile device and showed military and medical applications.

Recent work has explored different applications of multimodal interaction, including care-taking of older adults~\cite{10.1080/0144929x.2020.1851768,10.1007/978-3-319-69266-1_10}, photo editing~\cite{10.1145/2470654.2481301}, and digital virtual assistants~\cite{10.3115/v1/w14-4335}.
Researchers have also explored different devices and environments for multimodal interaction, including augmented reality~\cite{10.1145/3382507.3418850}, virtual reality~\cite{10.1145/1452392.1452444,10.1145/3491102.3502045}, wearables~\cite{10.1145/642611.642694}, and the Internet of Things~\cite{10.1145/3357251.3357581,10.1145/2858036.2858177,10.1145/3313831.3376427,10.1145/3332165.3347954}.

These projects have demonstrated the great potential of multimodal interaction systems.
However, multimodal systems still have limited adoption in the real world due to the development complexity they currently require.

\subsection{Graphical UI frameworks}

\label{sec:ui-history}

\mr{} is built on top of an existing graphical UI framework to provide a familiar development experience.
Model–view–controller (MVC)~\cite{10.5555/50757.50759} is the traditional basis of UI development frameworks and is used in frameworks such as Microsoft's Windows Forms~\cite{winforms}, and Apple's UIKit~\cite{uikit}.
The model stores data while the controller manages GUI input and updates the GUI view based on data changes.
Typically implemented in object-oriented programming languages, MVC can be compared to a shadow play, where objects (controllers) manipulate GUIs and data to maintain synchronization.
However, updating the model with alternative modalities, such as voice, is not feasible due to the strong entanglement between models and GUI updates.

Garnet~\cite{Myers1994, 10.1109/2.60882}, a user interface development environment introduced in the late 1980s, is another notable approach to GUI development.
Garnet introduced concepts like data binding, which allows the GUI to be updated automatically when the data changes.
It also tries abstracting the GUI state away from the presentation using interactors~\cite{Myers1990}.
While interactors allow the UI state to be rewired and thus to be updated using another modality like voice or gesture~\cite{Landay1993}, they do not enable manipulation of more abstract states (e.g., foods in a delivery order) that are not directly mapped to a single UI control.

Declarative UI frameworks, such as React~\cite{react}, Flutter~\cite{flutter}, and SwiftUI~\cite{swiftui}, are a more recent approach to UI development.
With declarative UI frameworks, programmers write functions to transform data into UI interfaces, and the system automatically manages updates.
To ease the management of states that may be updated by and reflected on multiple UI interfaces, centralized state management frameworks, such as Redux~\cite{redux}, Flux~\cite{flux}, and Pinia~\cite{pinia}, are often used together with these declarative UI frameworks.
They provide a single source of truth for the application state and allow state updates to be reflected across all presented UIs.
This approach can be likened to an overhead projector, where the centralized state represents the writing and the transform functions represent the lens projecting the UI to the user.
While this approach improves separation and UI updating, it sacrifices the object-oriented nature of the data model.
This centralized state works well with button pushes but comes short in dynamically composing multiple actions to support rich multimodal commands.

\mr{} reintroduces object-orientedness to centralized state management systems by representing the state as a sum of all class instances in the memory.
Developers can declare classes and describe actions as member functions of the classes.
\mr{} captures all instantiated classes and stores them in a central state.
This more modularized model is analogous to actors (class instances) in a movie set, with views (UI components) acting as cameras capturing different angles of the centralized state.
In this way, \mr{} enables rich action composition through type-checked function calls.
Furthermore, developers can tag specific cameras to point at certain objects, enabling automatic UI updates from state changes.
These features allow \mr{} apps to easily support the compositionality of multimodal input and enable the interleaving of multimodal input with other graphical UI actions.

\subsection{Voice UI frameworks}

\begin{figure*} 
    \centering
    \includegraphics{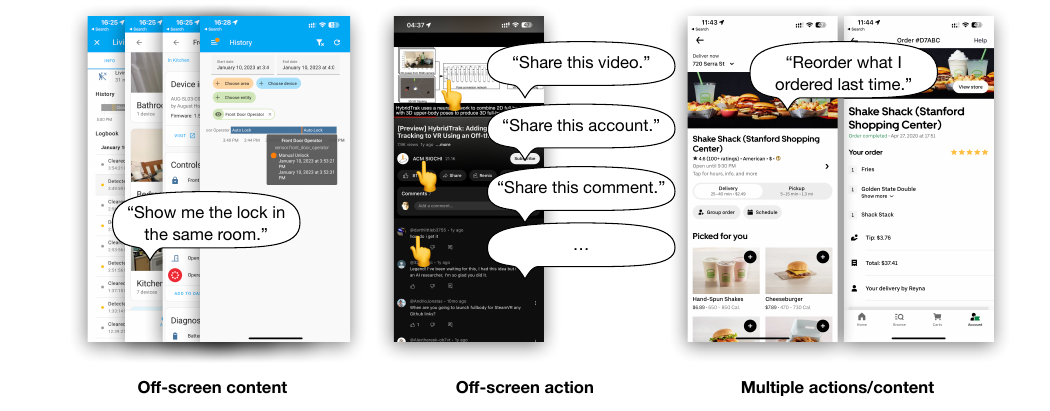}
    \Description{
        The figure shows three scenarios targeted by \mr{}.
        (left) Off-screen content:
        In a smart home app, the user views the history page of a motion sensor.
        They say, ``Show me the lock in the same room,'', and the app shows the lock operation history in the same room.
        (middle) Off-screen actions:
        In a video-playing app, the user says ``Share this video'', ``Share this account'', and ``Share this comment'' to share the video, the account, and the comment.
        (right) Multiple actions/content:
        In a restaurant app, the user says, ``Reorder what I ordered last time'' to reorder the same food items from the last order.
    }
    \caption{\mr{}'s targeted interaction scenarios.}
    \label{fig:interaction}
\end{figure*}

Commercial voice or chatbot frameworks, such as Amazon Lex, Google Dialogflow, and Rasa, are designed to handle natural language understanding and generation.
These frameworks allow developers to define intents and entities and then train the model to recognize the intents and entities from the user's input.
In this context, intents refer to categories of the user's action, such as making a reservation or asking for weather information, and one action can only be mapped to one intent.
Intents are usually mapped to different programming implementations to handle commands in the corresponding intent categories. 
These frameworks require a complete redevelopment of an application to support voice-only input.
Frameworks such as Alexa Skills Kit and Google Actions allow developers to extend existing applications to support voice input.
However, these still require manual work to build functions only for voice, and the visual UI updates are limited to simple text and a few pre-defined UI elements.
Additionally, the one-intent-one-implementation nature of the intent-based architecture limits the compositionality of the voice commands.

Research-focused voice/natural language frameworks, such as Genie~\cite{10.1145/3314221.3314594,10.1145/3340531.3411974} and other semantic parsers~\cite{sempre,10.1162/tacl_a_00190}, are designed to
support better compositionality of voice commands.
However, given that today's app development is primarily geared toward mobile and graphical interfaces, these frameworks require extra work from the developer and do not support multimodal features.
\mr{} improves this experience by integrating the development of voice and graphical UIs, allowing developers to extensively reuse existing code and support multimodal interactions.

\subsection{Multimodal Interaction Frameworks}

Prior work has also proposed multimodal interaction frameworks that allow developers to build multimodal applications.
One of the earliest works is presented by \citet{10.1109/38.773958}.
It includes ideas like forming the user's voice command as a function call and using the user's touch point as a parameter to the function call.
Later, researchers created standards~\cite{10.1007/s12193-013-0120-5, mmi-arch} and frameworks~\cite{10.1145/1358628.1358881, 10.1145/1570433.1570476, 10.1145/1979742.1979703, 10.1080/088395199117504, 10.1145/2070481.2070500, 10.48550/arxiv.2103.15975, 10.1109/vrw55335.2022.00018} to help developers build apps that can handle multiple inputs across different devices.
Although these frameworks provide scaffolding for developers to build multimodal applications, they mostly treated voice as an event source that can trigger functions the developer has to explicitly implement for voice.
Developers also have to manually update the UI to reflect the result of the voice command.
This manual process limits voice commands to simple single-action commands and makes it difficult for developers to build richer multimodal applications.

Recently, there are research projects on generating voice commands by learning from demonstration~\cite{10.1145/3453483.3454046, 10.1145/3379337.3415848,10.1145/3210240.3210339}, extracting from graphical user interfaces with large language models~\cite{garg2023,10.48550/arxiv.2209.08655}, or building multimodal applications using existing voice skills~\cite{10.1145/3379337.3415841}.
The first approach still requires developers to manually create demonstrations for each action and limits the compositionality of the voice commands.
The second approach is useful for accessibility purposes, but it relies on the features being easily extractable from the GUI\@.
It is uncertain how well the first two approaches can generalize to more complex UI tasks that require multiple UI actions.
The third approach is constrained by what is provided by the voice skills and, traditionally, these have been limited due to the added development effort.

In comparison, \mr{} leverages the existing GUI development workflow and requires only minimal annotations to the code to generate multimodal applications.
Having access to the full object-oriented state programming codebase, \mr{} can handle the natural complexity of multimodal input, compose the right series of function calls, and update the UI to reflect the result automatically.

\section{System Design}

In this section, we first define the design goals of the framework.
Then, we describe the theory of operation that addresses the design goals.
Finally, we discuss the implementation of the system components and workflow.

\subsection{Design goals}

Our design goals include aspects of the interaction design of \mr{} apps as well as the design of the framework itself.

\label{sec:targeted-interactions}

\subsubsection{Interaction Design}
\mr{} is primarily designed to enhance user interaction with mobile applications, but the concept should also apply to apps on other platforms.
Today, mobile applications are well-optimized for touch and graphical interactions.
Users can use the graphical interface to see content on the screen and use touch to access actions on the screen.
To further enhance the user's performance and reduce cognitive load, \mr{} focuses on supporting interactions that often involve touch actions used together with a voice command.

Here is a series of example commands in interactions with a food ordering app between user A and their friend user B:
\begin{enumerate}
\item[\textbf{C1}] A knows what they want, so A says, ``Show me what I ordered last week from McDonald's.'' The app responds with the order history.
\item[\textbf{C2}] A wants to add a previously ordered food into the cart (not available on UI). A says, ``Order this hamburger,'' with a tap on the ``Big Mac'' entry in the order history, and the app adds a ``Big Mac'' to the shopping cart.
\item[\textbf{C3}] B wants to order something different, so they tap on the restaurant to view the menu.
\item[\textbf{C4}] B doesn't like beef, so they say, ``Show me food without beef,'' the app displays options accordingly.
\item[\textbf{C5}] B says, ``Order a meal with this sandwich,'' with a tap on the ``McChicken sandwich,'' the app adds a McChicken meal to the cart.
\end{enumerate}

This interaction demonstrates the power of these multimodal commands where voice and touch are used interchangeably or in conjunction.
These commands can be categorized into three interaction design goals (see \autoref{fig:interaction}):

\aptLtoX[graphic=no,type=html]{
\begin{enumerate}
    \item[\textbf{I1}] \textbf{Access off-screen content} (C1, C4):\label{i:access-off-screen-content}
    For example, the user is looking at the smart home app and notices abnormal motion on the smart home app's living room security camera.
    So they can talk to the smart home app, ``Show me the lock status history in the same room of this device.''
    This interaction usually requires multiple GUI navigation steps (go to the device's room page, navigate to the door lock section, check door lock history) to access the content.
    \item[\textbf{I2}] \textbf{Access off-screen actions} (C2):\label{i:access-off-screen-actions}
    For example, the user says, ``Share this creator/comment'' while watching a YouTube video.
    Some actions are hidden behind a menu or a button, and some are not accessible at all on mobile devices.
    \item[\textbf{I3}] \textbf{Combine multiple actions/content} (C5):\label{i:combine-multiple-actions-content}
    For example, the user says, ``Order what I ordered last time'' while looking at a food delivery app.
    This usually requires the user to go back and forth between an order detail page and a menu page.
\end{enumerate}
}{
\begin{enumerate}[label=\textbf{I\arabic*}]
    \item \textbf{Access off-screen content} (C1, C4):\label{i:access-off-screen-content}
    For example, the user is looking at the smart home app and notices abnormal motion on the smart home app's living room security camera.
    So they can talk to the smart home app, ``Show me the lock status history in the same room of this device.''
    This interaction usually requires multiple GUI navigation steps (go to the device's room page, navigate to the door lock section, check door lock history) to access the content.
    \item \textbf{Access off-screen actions} (C2):\label{i:access-off-screen-actions}
    For example, the user says, ``Share this creator/comment'' while watching a YouTube video.
    Some actions are hidden behind a menu or a button, and some are not accessible at all on mobile devices.
    \item \textbf{Combine multiple actions/content} (C5):\label{i:combine-multiple-actions-content}
    For example, the user says, ``Order what I ordered last time'' while looking at a food delivery app.
    This usually requires the user to go back and forth between an order detail page and a menu page.
\end{enumerate}
}

The common theme among these interactions is that they require the multimodal interaction framework to understand the content and actions available in the app.
For content, the framework needs to know what is the content on the screen, how to access it, and how to represent returned content (from user-initiated commands) to show the retrieved content.
For actions, the framework needs to know the list of available actions and how to render changes on the user interface after the action is triggered.
Finally, the framework needs to translate the user's intent, which may be rich, into possibly a series of actions and content displays.

With \mr{}, apps will have a microphone button on the screen.
When the user taps on the button, the user can say their command and refer to the content on the screen by tapping it.
The app will then parse the voice command and touch input and execute the corresponding actions to help the user with the scenarios described above.

\subsubsection{Framework Design}

\begin{figure*}
    \centering
    \includegraphics[width=\linewidth]{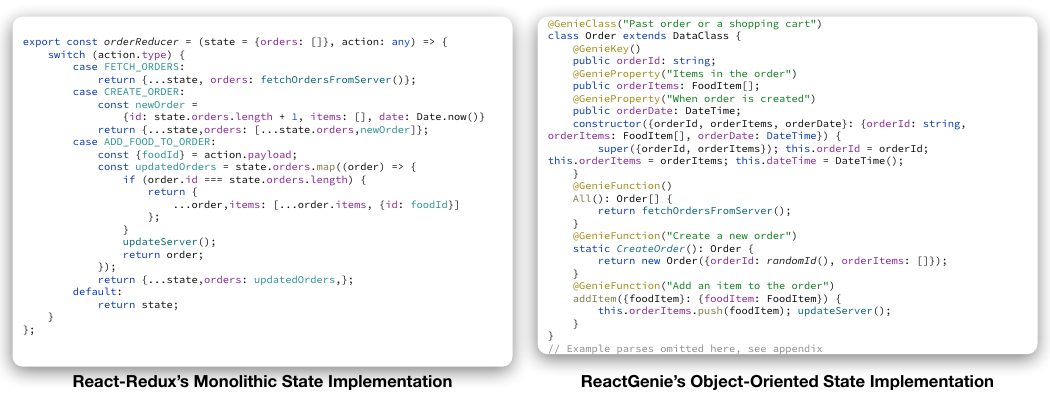}
    \caption{
        A comparison between state code in React-Redux and in \mr{}.
        (Left) Part of an example state code in Redux.
        Data is stored in the state variable, and the state can be mutated by the actions defined.
        These actions (stored in a \texttt{Reducer}) do not have explicit types, and they directly manipulate the state, so no return values are defined. 
        Note that the return values of case statements in a \texttt{Reducer} indicate the new state variable after the state changes; actions do not have return values.
        Due to its monolithic design, it is hard to compose functions together to achieve some multimodal actions.
        \enspace
        (Right) Part of an example state code in \mr{}.
        Automatically managed by \mr{}, the state is composed of all the instantiated objects' \texttt{DataClass}es.
        Actions in the state code are defined as methods of the class. All the methods have explicit parameter types and return types.
        These functions can be composed together to achieve multimodal actions.
    }
    \Description{
        The figures show two snippets of code, one from Redux and one from \mr{}.
    }
    \label{fig:state-code}
\end{figure*}

To translate the content and actions in the user's rich multimodal intent, \mr{} needs to obtain information about the app's capabilities from the code.
The design goal for \mr{} is to do this in a way that causes minimal disruption for the application developer.

Without a proper framework for multimodal apps, the developer must design their own mechanisms to handle voice, handle the complexity of multimodal commands, and maintain control of the app's behavior.
Our goals for the \mr{} framework include handling these issues:
\aptLtoX[graphic=no,type=html]{
\begin{enumerate}
    \item[\textbf{F1}] \textbf{Maintain the expressiveness} of the developer for their GUI appearances and specific app functionality.
    \label{f:allow-full-control}
    \item[\textbf{F2}] \textbf{Reduce development cost} by maximizing the reuse of existing GUI code and hiding the complexity of handling multimodal commands from developers.
    \label{f:maximize-reuse}
    \item[\textbf{F3}] \textbf{Ease the learning curve} by providing a similar programming experience to existing GUI frameworks.
    \label{f:ease-learning-curve}
\end{enumerate}
}{
\begin{enumerate}[label=\textbf{F\arabic*}]
    \item \textbf{Maintain the expressiveness} of the developer for their GUI appearances and specific app functionality.
    \label{f:allow-full-control}
    \item \textbf{Reduce development cost} by maximizing the reuse of existing GUI code and hiding the complexity of handling multimodal commands from developers.
    \label{f:maximize-reuse}
    \item \textbf{Ease the learning curve} by providing a similar programming experience to existing GUI frameworks.
    \label{f:ease-learning-curve}
\end{enumerate}
}
\subsection{Theory of Operation}

\begin{figure*}
    \centering
    \includegraphics[width=\linewidth]{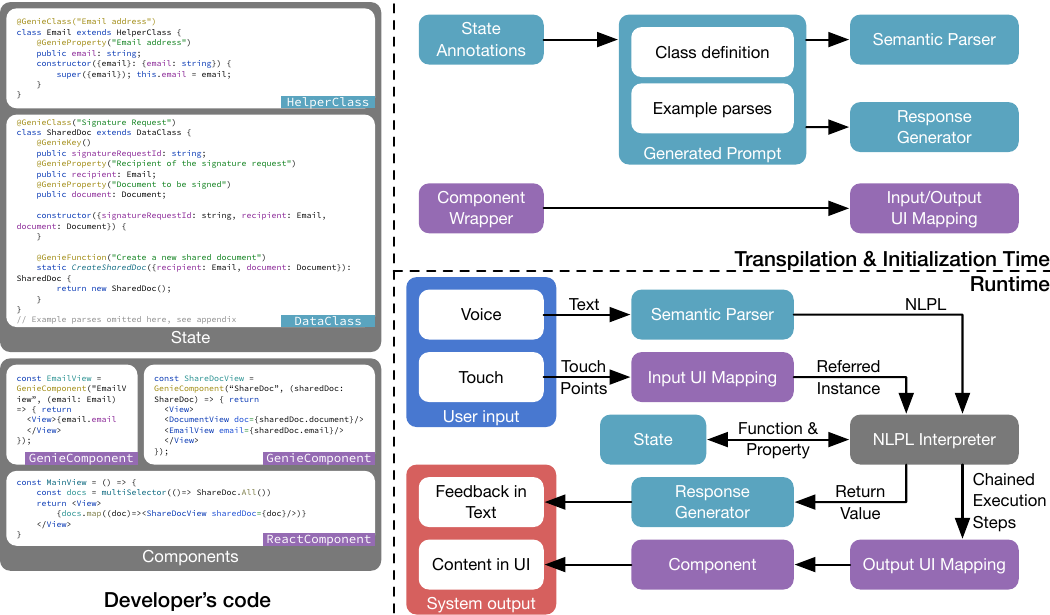}
    \caption{
        Overview of the \mr{} system:
        (Left) Developers write object-oriented state code for programming content and actions and define the UI as cascading components.
        (Right top)
        \mr{} operates at transpilation and initialization time to generate runtime modules.
        (Right bottom)
        Developer modules, generated modules, and \mr{} modules come together to process rich multimodal commands from the user.
        This workflow is similar to regular GUI development, maximizes code reuse, and allows full control of the app behavior.
    }
    \Description{
        The figure shows the overview of the \mr{} system.
        On the left is the developer's code.
        The code includes the object-oriented state code (top left) and the UI code (bottom left).
        On the left is what \mr{} does to help with multimodal interactions.
        On the top left it describes the transpilation and initialization process.
        On the bottom left it describes the runtime process.
    }
    \label{fig:overview}
\end{figure*}

\mr{} presents an object-oriented state programming model to the developer.
\textit{State code}, in the context of GUI development, refers to the part of the application that manages the data and logic that determine the state of the user interface.
The global state store in Redux 
is typically represented as a tree of stored variables and the associated functions to transform them.
The concept of state in declarative UIs is similar to the model in the model-view-controller (MVC) programming models.

As mentioned in \autoref{sec:ui-history}, in frameworks like React, UI development is moving towards separating the UI from the state.
Developers define functions (\textit{components}) that convert the state into UI.
Each UI component receives part of the state and renders the UI in an HTML-like format.
Components can sometimes render a part of their state using another component in a compositional way (e.g., a restaurant menu component can use menu item components to render each food item, and the restaurant menu components host a list to organize the menu item components).
Therefore, there is a unidirectional data flow from the state to the UI\@ so that the state update is automatically reflected on the UI without any additional code from the developers.
In comparison, in a typical MVC paradigm, the controller simultaneously updates the model (state) and the UI to keep them in sync.
The unidirectional data flow feature in declarative UIs (compared to MVC) allows data to be updated outside of GUI input because data updates are no longer entangled with UI update code in MVC's controller code.
This feature allows \mr{} to use multimodal commands to change the same state and update the UI accordingly, maximizing the reuse of existing GUI code (\ref{f:maximize-reuse}).

However, in these existing declarative UI frameworks, the state is represented by a single global data store.
This data store contains a tree-like data structure with all the content and status of the app (state) and a list of functions to manipulate the state data.
The functions usually contain parameters for indexing into the state object and parameters to further specify the actions.
For example, an \texttt{ADD\_FOOD\_TO\_ORDER} action needs to have the parameters of \texttt{food\_id}, \texttt{order\_id}, and \texttt{quantity}.
This works well for GUI design, as developers can implement an action for each button press, and each button in the GUI stores the corresponding ID that it needs to call when pressed.
However, this makes it difficult to handle typical multimodal commands, which require composing multiple actions (\ref{i:combine-multiple-actions-content}).
To translate a user command of ``Add two hamburgers to a new order'', an example translated program in React-Redux would be:
\begin{verbatim}
// create the order
dispatch(CREATE_ORDER())
// find the id of the created order
const order_id = store.orders.last().id
// find the food id of hamburger
const food_id = store.foods.
  filter(food => food.name === 'hamburger')[0].id
// add food to order
dispatch(ADD_FOOD_TO_ORDER(order_id, food_id, 2))
\end{verbatim}
This process involves the creation of multiple intermediary variables and queries to the state object.
When the neural semantic parser generates code for this process, we have found that LLMs will often generate a line of code referencing an intermediary variable that has not been declared before, causing an error in response to the user's query.
The problem is that even if we have already created a new order, we would have to retrieve the order ID from the state object, save it in an intermediary variable, and feed that ID into the imperative style actions.
This issue is solved in \mr{} by applying concepts of object-oriented programming where an order can be represented by a separate object that has both the data and all of the relevant actions (e.g., adding a food item to the order) associated with the object, so a retrieved order can directly be used to call the add food item action.

So, \mr{} introduces the object-oriented programming model to componentize the state of a declarative UI app.
In componentizing the state object, developers implement smaller objects containing its relevant content stored in properties and actions defined as methods.
This componentized design allows \mr{} to translate a typical multimodal command into a single statement with method chaining.
Using the example above, the user's command can be translated as: \texttt{Order.CreateOrder().addFoodItem(foodItem: FoodItem.All().matching(field: .name, value: "hamburger"), quantity: 2}.
With \mr{}'s state abstraction, LLMs can generate code that directly calls \texttt{addFoodItem} after creating the new order.
These methods are also strictly typed, which helps the natural semantic parser develop the correct combination of methods with fewer runtime errors.
This succinctness improves the accuracy of the neural semantic parser that translates the user's natural language command to executable code.
Furthermore, as the GUI and voice modality share the same content and state, this representation supports interchangeability in user input modality for each part of the rich multimodal command.

In practice, to work with \mr{}'s object-oriented state abstraction, the developer identifies the user-accessible content (object or object properties) and actions (functions) with the \texttt{GenieProperty} and \texttt{GenieFunction} annotations, respectively, along with an example of how it may be referred to in English as shown in \autoref{fig:state-code} right.
The GUI is programmed using compositional components (similar to other declarative UI frameworks described above) rendered from the user-accessible state objects.
This allows the internal state to be rendered to the user in a GUI\@.
The high-level model of \mr{} resembles popular GUI development frameworks (React + Redux), which makes it easy for developers to learn and use (\ref{f:ease-learning-curve}).
\mr{} automatically handles the retrieval of content (objects/properties) off-screen (\ref{i:access-off-screen-content}), the execution of actions off-screen (\ref{i:access-off-screen-actions}), and combinations of the two (\ref{i:combine-multiple-actions-content}).
In this way, voice and multimodal commands will be handled by LLM-generated programs and touch commands can be handled in the traditional way where developers write programs to handle each individual user input event.

By reusing a programming model similar to existing UI frameworks, \mr{} allows developers to control the look and feel of their app to a similar degree as typical declarative UI frameworks.
Meanwhile, developers can also control what function the end-users can access in multimodal interactions through whether or not to add annotations.
Hence, developers have full control over the functionality of the app.
These designs help developers maintain full control of their app while enjoying the benefits multimodal interaction offers to end-users (\ref{f:allow-full-control}).

\subsection{The Developer's Programming Model}

\label{sec:developer-modules}
\begin{table*}[ht]
\caption{Annotations and Functions for programming \mr{}.}
\label{tab:annotations-functions} 
\begin{tabular}{lp{2cm}p{1.5cm}p{4cm}p{6.0cm}}
\toprule
Annotation/Function & Location & Type & Parameters & Purpose \\
\midrule
\texttt{GenieClass} & DataClass/ HelperClass & Annotation & \texttt{comment} (optional): describing the class's purpose. & Indicates the class to be user-accessible through multimodal interactions, essential for determining which parts of the code are relevant to the multimodal experience. \\
\texttt{GenieFunction} & DataClass/ HelperClass & Annotation & \texttt{comment} (optional): describing the function's purpose. & Tags a method to be exposed to multimodal interaction, marking this method as something the user may call directly. \\
\texttt{GenieProperty} & DataClass/ HelperClass & Annotation & \texttt{comment} (optional): describing the property's purpose. & Exposes the property for multimodal interaction, letting users interact with this property through multimodal commands. \\
\texttt{GenieKey} & DataClass & Annotation & None required. & Identifies a unique ID property that uniquely identifies an instance within a DataClass, crucial for managing and retrieving data instances. \\
\texttt{constructor} & DataClass/ HelperClass & Function & Varies based on class requirements. & Initializes a DataClass instance with required data, setting up the basic data structure of an object. \\
\midrule
\texttt{GenieComponent} & UI Component & Wrapper Function &  \multicolumn{1}{p{9pc}}{\texttt{target}: DataClass or HelperClass to be displayed.\hfill\break \texttt{component}: The UI component to be wrapped.} & Wraps a UI component and maps it to a DataClass or HelperClass instance, telling how to present data/state visually and enable touch references in multimodal interactions. \\
\bottomrule
\end{tabular}
\end{table*}


From the developers' point of view, using \mr{} is similar to any declarative UI framework.
They need to implement the state code that provides the content and actions supported in the app.
They also need to specify the UI components that translates current state classes into UI interfaces that the user can see and manipulate.
\autoref{tab:annotations-functions} provides a list of functions and annotations that developers need to provide for \mr{} (see \autoref{fig:overview} left for an example).

In \autoref{sec:system-workflow}, we will describe in detail the right side of  \autoref{fig:overview}, which is how the \mr{} system uses the developer-supplied code described in this section, (1) transpiles (source-to-source compiles) it into \mr{} modules (Right top), and (2) uses the generated \mr{} modules to handle end-users' input (Right bottom).

\subsubsection{State Code}

Developers provide the content and actions in an app through \mr{}'s object-oriented state model, implemented in TypeScript\footnote{\href{https://web.archive.org/web/20240228110208/https://www.typescriptlang.org/}{https://www.typescriptlang.org/}}.
As with all object-oriented programming models, \mr{}'s state code includes the definition and implementation of classes.
Classes have declarations, methods, and properties, which can be labeled as \texttt{GenieClass}, \texttt{GenieFunction}, and \texttt{GenieProperty}.
These annotations in the state code (or \textit{State Annotations} for short) indicate that they are user-accessible via multimodal commands.
All \mr{} annotations have an optional parameter that denotes the purpose of that class/function/property, similar to comments in code.
These code annotations or decorators in TypeScript are tags written before the class, method, and property declarations.
This allows the relevant annotation code to be executed at initialization time to transform the capabilities of the annotated classes or methods.

All states that are accessible via multimodal interactions must be declared as instances of the \texttt{DataClass} and \texttt{HelperClass} provided by \mr{}.
A \texttt{DataClass} stores the app's data, and a \texttt{HelperClass} provides definitions to ease the user's interaction with the data. The properties of a \texttt{DataClass} can be of TypeScript primitive types, a \texttt{DataClass}, or a \texttt{HelperClass}.

\autoref{fig:overview} left shows an example of a document signing app; examples of each class are shown in the top left.
As shown in the figure, the \texttt{SharedDoc} \texttt{DataClass} tracks the lifecycle of a document's signature request.
An \texttt{Email} class is a \texttt{HelperClass} that helps manage users' email addresses.
\mr{} also has system-provided \texttt{HelperClass} such as \texttt{DateTime} and \texttt{TimeDelta} to help developers manage date, time, and period of time.
The introduction of the \texttt{HelperClass} class not only allows developers to define helper functions but also helps with type-checking, which is useful for developers to write less buggy code and for \mr{}'s semantic parser to generate more correct \mrd{}.

All \texttt{DataClass}es start with the class declaration, which begins with the \texttt{GenieClass} annotation and then the class name and a required inheritance of \texttt{DataClass}.
See the first two lines of \texttt{ShareDoc} in \autoref{fig:overview}.
Developers need to implement one method and one property:
\begin{enumerate}
    \item \texttt{constructor} method:
    The constructor of the class initializes the instance with all the required data.
    \item \texttt{id} property:
    A unique identifier of the instance, annotated by the \texttt{GenieKey}.
\end{enumerate}

Developers can also implement additional functions and properties to complete the \texttt{DataClass}.
If developers want the user to be able to use the functionality directly, they need to add the \texttt{GenieFunction} and \texttt{GenieProperty} annotation.
An example of that can be found in \autoref{fig:overview} left (see \texttt{document} and the \texttt{CreateSharedDoc} in \texttt{SharedDoc}).
Sometimes, developers may want to implement internal housekeeping functions, such as clearing the application cache or managing internal database transactions.
Those functions/properties should not be annotated and will not be called by \mr{}.


The \texttt{HelperClass} allows developers to define new types that can be used in the \texttt{DataClass}.
A specific example is the \mr{}-supplied \texttt{DateTime} \texttt{HelperClass}.
It can support operations like offsetting the time by a certain amount or setting the year/month/day/hour/minute/second/day of the week to a certain value.
It allows commands such as ``last Thursday'' to be translated to ``\texttt{DateTime.Current.offset(week: -1).set(weekOfTheDay: 4)}''.
The developer can define other \texttt{HelperClass} instances to support more complex operations such as length unit conversion.
Similarly, the \texttt{HelperClass} needs to have a \texttt{GenieClass} annotation and needs to inherit from \texttt{HelperClass}.
The \texttt{HelperClass} only requires a \texttt{constructor} method that takes the data as input and initializes the instance.
\mr{} will also generate a \texttt{Current} property for the \texttt{HelperClass} that returns the instance that is being referred to by the user.
\mr{} does not keep track of the instances of \texttt{HelperClass} separately in memory, but instead, they will be part of the \texttt{DataClass} that uses them.

For both \texttt{DataClass} and \texttt{HelperClass}, the developer can define a \texttt{description} method to customize the string representation of the instance for response generation.
By default, \mr{} will generate a JSON-like representation using all the instance's properties.

Finally, developers need to provide example parses for the \texttt{DataClass} and \texttt{HelperClass}.
\textit{Example parses} or \textit{few-shot examples} are pairs of expected end-user voice command examples and the corresponding translated \mrd{}.
Developers provide them as few-shot examples for \mr{}'s neural semantic parser.
These examples are helpful for the parser to learn about what the app is about and the syntax of \mrd{}.
In practice, around 10 example parses are sufficient.
Developers do not have to cover all use cases, and \mr{}'s semantic parser can automatically generalize to most of the app's features and user expressions.

\subsubsection{Built-in Dataclass Methods}

\mr{} automatically generate three methods for each \texttt{DataClass}:
\begin{enumerate}
    \item \texttt{All} method:
    For voice inputs, the user needs to refer to all or a select set of instances of certain objects.
    A static \texttt{All} method is provided that returns an array of all the instances of the DataClass.
There are built-in functions that support filtering array elements based on value or value ranges, sorting based on an order on a property, and extract an element by its index.
    \item \texttt{Get} method:
    A static method that takes an \texttt{id} as input and returns the instance with the corresponding \texttt{id}.
    \item \texttt{Current} property:
    A static method that returns the instance that is being referred to by the user.
    This is automatically annotated with \texttt{GenieProperty}.
\end{enumerate}

Like many of the common state management frameworks, when any of the properties of a \texttt{DataClass} instance are changed, UI components on screen that refer to that property will be automatically re-rendered with the most up-to-date data.
This ensures that the UI is always in sync with what is being represented in the state.

\mr{} automatically maintains the instances of \texttt{DataClass} in memory to form the app's state.
\texttt{DataClass} can be backed up remotely, which is common in modern app development.

\subsubsection{Component/UI Code}
\mr{} developers need to define the GUI (as shown in the three code boxes in the bottom left of \autoref{fig:overview}) as a set of functional components\footnote{\href{https://web.archive.org/web/20240229091735/https://react.dev/learn/your-first-component}{https://react.dev/learn/your-first-component}}, similar with React.
These components can refer to each other to facilitate reuse.
It is common for every single instance of a \texttt{DataClass} or \texttt{HelperClass} to be represented by a component.
For example, \texttt{EmailView} represents an \texttt{Email} instance, while \texttt{SharedDocView} represents a \texttt{SharedDoc} instance.
Therefore, \mr{} introduces a special wrapper function (or \textit{Component Wrapper} for short) called \texttt{GenieComponent} to show that explicit mapping.
Instead of the arbitrary parameters of a normal functional component, components wrapped by \texttt{GenieComponent} take a \texttt{DataClass} or \texttt{HelperClass} instance as input.
\texttt{GenieComponent} allows the \mr{} runtime to understand which component is mapped to which instance in memory to facilitate reference by touch.
It also allows \mr{} to render the result of the user's request using the developer-defined component.
While defining \texttt{GenieComponent}, the developer can also specify an optional title and priority (both can be a method of the state instance) for the interface, which is relevant for choosing the interface to render multimodal command responses.

\subsection{\mrd{}}

We use a neural network to translate natural language commands into \mrd{},
a domain-specific language (DSL) we created. 
We feed the large-language model the automatically-extracted developer's class skeleton (only declaration and type information, no implementations to save tokens, and reduce distractions), developer-supplied few-shot examples, and the user's current voice command and ask it to generate \mrd{} code for understanding the user's intention.

We do not generate JavaScript directly because the expressiveness of JavaScript may cause unintended changes to the app's behavior (contradicting~\ref{f:allow-full-control}).
We cannot use a simple intent classifier, such as the one used by traditional voice assistants, because of the complexity of our user's multimodal commands.
The \textit{DSL interpreter} module runs the generated \mrd{} code and calls the corresponding methods in the developer's state definition code (called \textit{state code} in the following sections).
This also allows \mr{} to handle users' verbal references for simultaneous touch input through special reference functions (see \autoref{sec:transpilation}).

\label{sec:dsl-design}

The \mrd{} is designed to meet these language design goals:
\aptLtoX[graphic=no,type=html]{
\begin{enumerate}
\item[L1] \textit{Easy to generate}: \label{l:generate}
LLMs can generate syntactically correct \mrd{} code.
\item[L2] \textit{Robust against generation errors}: \label{l:robust}
LLMs can generate semantically correct \mrd{} code.
\item[L3] \textit{Able to express multimodal commands}: \label{l:expressiveness}
\mrd{} can express diverse multimodal commands.
\end{enumerate}
}{
\begin{enumerate}[label=\textbf{L\arabic*}]
\item \textit{Easy to generate}: \label{l:generate}
LLMs can generate syntactically correct \mrd{} code.
\item \textit{Robust against generation errors}: \label{l:robust}
LLMs can generate semantically correct \mrd{} code.
\item \textit{Able to express multimodal commands}: \label{l:expressiveness}
\mrd{} can express diverse multimodal commands.
\end{enumerate}
}

Therefore, because of~\ref{l:generate}, \mrd{} has to be in a form similar to existing programming languages that LLMs are trained on.
To help with~\ref{l:robust}, we also want \mrd{} to be strongly typed.
We tried a syntax similar to TypeScript, but we noticed the LLM-based semantic parser occasionally generated the correct parameters but in the wrong order.
Therefore, we decided to use a syntax similar to Swift, a strongly typed language that requires parameter names to be specified in the function call.

To support the expressiveness of human languages~\ref{l:expressiveness}, a simple data formatting language such as JSON is not enough.
JSON is good at representing structured data and is sometimes used to represent simple intent with a single function call and parameters.
However, the user's commands can consist of multiple method invocations and represent complex logic flow, so we looked at the language structures of what people may say to a \mr{} app.
When interacting with a virtual assistant, people typically utter an imperative or interrogative sentence.

An imperative sentence must have a verb and an object.
Some more complex imperative sentences can have object modifiers and verb modifiers.
For example, ``[Change the background color](verb) of [all the yellow](object modifier) [textboxes](object) [to orange](verb modifier)''.
Objects can be translated to function calls to retrieve the corresponding objects, e.g., ``this food'' \texttt{-> Food.Current()} and ``textboxes'' \texttt{-> TextBox.All()}.
Object modifiers can be translated to SQL-like operations, e.g., ``food with a rating above 3'' \texttt{-> Food.All().between(field: .rating, from: 3)}, ``all the yellow textboxes'' \texttt{-> TextBox.All().equals(field: .color, value: "yellow")}.
Note that we did not use SQL syntax because SQL does not have easy support for calling functions of objects, so it would have trouble translating verbs/actions.
Verb and verb modifiers are translated to function calls.
For example, changing the background color to orange would be \texttt{.setBackgroundColor(color: "orange")}.

We also avoid supporting lambda expressions, variable declaration, and control flow statements to avoid reference errors to increase robustness (\ref{l:robust}) in translation and reduce complexity.
For lambda expressions, \mrd{} automatically distributes function calls to individual elements of objects to support plural objects: ``make all textboxes orange'' \texttt{-> TextBox.All().setBackgroundColor(color: "orange")}.
For variable declaration, we use method chaining to avoid reference errors.
This is because, in early testing, we found that OpenAI Codex frequently refers to undeclared variables.
Prior work on building a target language for a neural semantic parser also observed similar problems~\cite{10.48550/arxiv.2203.12751,10.1145/3038912.3052562}.
For control flow statements, we can use function call distributions for for-loops and use the SQL-like syntax described above for if-statements.

In an interrogative sentence, the verb matters less (typically, the verb is just ``is''), and there are still object and object modifiers.
We can translate the object and the object modifiers similarly:
``When is [the last time](subject) [I ordered from this restaurant](subject modifier)'' \texttt{-> Order.All().equals(field: .restaurant, value: Restaurant.Current()).sort(field: .date, ascending: false)[0].date}.

Human languages are flexible and do not always have to follow the exact grammar.
Still, an LLM can automatically compose the \mrd{} features above to accommodate the user's request as long as the features are in the developer's program.
The full grammar of \mrd{} is listed in Appendix~\ref{sec:grammar}.
We implemented the \textit{DSL interpreter} module using the peggy\footnote{\href{https://web.archive.org/web/20240215190544/https://peggyjs.org/}{https://peggyjs.org/}} parser generator.

\subsection{System Workflow}
\label{sec:system-workflow}

As shown in \autoref{fig:overview} right, \mr{} provides libraries that automatically perform actions during two phases of the development and usage process: 1) transpilation and initialization time and 2) runtime.

At transpilation and initialization time (\autoref{fig:overview} right top), \mr{} uses the annotations in the state code to generate LLM prompts containing class definitions and example parses for the semantic parser and response generator modules (\autoref{sec:transpile-state-code}).
\mr{} also reads the component wrapper code to determine the mapping between the components and the state objects for the input and output UI mapping modules (\autoref{sec:transpile-ui-code}).

At runtime (\autoref{fig:overview} right bottom), \mr{} processes the user's multimodal voice and touch input.
The voice part is translated to \mrd{} using the semantic parser, and the touch points are translated to the referred state instances using the input UI mapping module.
Knowing the \mrd{} and the referred state instances, the \mrd{} interpreter can call the relevant methods and properties in the developer's state code to achieve the user's request.
The \mrd{} interpreter records the final return value and the intermediary execution steps from executing each part of the composed method-chaining statement.
\mr{} further uses the final return value to generate text feedback for the user.
It also uses the execution steps to graphically present the answer to a query-type request or the effect of an action-type request.

Note that \textit{Transpilation} is a source-to-source translation process from the TypeScript the developer writes to Javascript that the machine executes.
Typically, a TypeScript app is transpiled to JavaScript to run in a mobile app or a browser.
However, during the transpilation process, the metadata, like typing and function parameter names, are removed.
The metadata lost in transpilation is required by \mr{} to create a large language model-based semantic parser/response generator and the UI mapping.
Therefore, we built our transpliation plugins to use the developer's code before transpilation to generate the modules that \mr{} uses at runtime.

\subsubsection{Transpilation and Initialization for State Code}

\label{sec:transpile-state-code}
\label{sec:generated-modules}

\mr{} uses a custom transpiler plugin that generates extra metadata for \texttt{@GenieProperty}, and \texttt{@GenieFunction} of \texttt{DataClass} and \texttt{HelperClass}.

We use in-context learning to implement the semantic parser and the response generator. 
During initialization of the app, \mr{} will load injected metadata from \textit{state classes} (classes in state code) to generate a base prompt shared by both the semantic parser and the response generator\@.
LLMs work by generating text continuations given a paragraph of previous text.
The provided previous text is often called the \textit{prompt}.
By controlling the prompt, we change the information the LLM has access to and guide the LLM to do what we want (generate the corresponding \mrd{} of the user's command).
\mr{}'s {\em generated prompt} contains two parts:
1) The \textit{class definitions} contain all the \texttt{DataClass} and \texttt{HelperClass} method and property definitions with the implementation stripped out.
It is rendered in a format similar to Swift syntax.
2) The \textit{example parses} provided by the developer are also included as few-shot examples.

The user input is then appended to the generated prompt and used in the LLM-based semantic parser for \mrd{} translation.
The \textit{response generator} prompts the LLM with the generated prompt, the user input, the parsed \mrd{}, and the \texttt{description} of the return value from the execution of \mrd{} to produce a short text response.

We built the semantic parser using the OpenAI Codex model \texttt{code-davanci-2} and the response generator using the OpenAI GPT 3.5 model \texttt{text-davanci-3}.

\subsubsection{Transpilation and Initialization for UI Code}
\label{sec:transpile-ui-code}

\label{sec:transpilation}

At initialization time, we also process \texttt{GenieComponent} functions to save a mapping between the \texttt{GenieComponent} and the \texttt{GenieClass} that they are representing.
We generate \textit{input and output UI mapping} modules from this information.
We monitor the bounding box of all \texttt{GenieComponent}s for input mapping.
When the user touches the screen while expressing a multimodal command, \mr{} will use the bounding box information to determine which component the user is pointing to.

It is common for multiple UI components to cover the area where the user taps on the screen.
For example, in \autoref{fig:teaser}, all the \texttt{FoodThumbnail} components overlap with the \texttt{OrderItemView} components.
\mr{} allows the user to use their voice to disambiguate the reference:
If the user mentions food, such as ``this food'' (\texttt{FoodItem.Current()}), or actions that can only be done with food, like ``what is the price for this'' (\texttt{FoodItem.Current().price}), \mr{} will use the \texttt{FoodItem} object and vice versa.
In the special case where multiple components of the same type cover the tapped area, \mr{} uses the one with the smallest bounding box.

Another common scenario is that if one object is clearly in the ``foreground'' of the graphical UI, the user may naturally refer to it as ``this'' without explicitly specifying the component via touch.
So, when the user refers to a state class and either there is no touch point, or the touch point does not match any component representing that class, \mr{} will use the largest component on the screen representing that class as the reference.

We also use \texttt{GenieComponent} to generate \textit{output UI mapping} modules.
We gather all the \texttt{GenieComponent}s with supplied priority and title and group them by the state class they represent.
When the result of the executed \mrd{} is a state class instance, \mr{} enumerates through all the \texttt{GenieComponent}s representing that class and renders the one with the highest priority.

There are two types of execution results.
The first type is for query-type requests where the translated \mrd{} returns an instance that can be rendered by a \texttt{GenieComponent}.
This is common when the user asks to either retrieve some data ``what are my most recent orders from this restaurant?'' or to perform some action with a clear result ``what vegetarian food does this restaurant offer?''.
In that case, rendering the result on the screen would be intuitive.
So, when the return value can be represented by a \texttt{GenieComponent}, \mr{} will always find the highest priority \texttt{GenieComponent} and render it.

The second type is for action-type requests where the translated \mrd{} returns a value that cannot be rendered by a \texttt{GenieComponent}.
For example, if the user asks to ``add a hamburger to the cart'' (\mrd{}: \texttt{Order.GetActiveCart().addItem([Food.Named("hamburger")])}, it would return \texttt{void} which cannot be rendered.
For these actions, the return value is less important, and the user is more interested in the side effects of the action (e.g., the cart has been updated).
In that case, \mr{} will back trace the chained execution steps and find the last renderable result is the return value of \texttt{Order.GetActiveCart()} (i.e., an instance of a cart).
\mr{} will also check all the currently visible components to see if there is any that already represent the same instance of that cart on screen.
\mr{} would only render this result if the current page has no component representing the same instance.
For example, when the user is already on a restaurant page where they can see an indicator of the number of items in the cart (e.g., the cart icon with a counter also represents the cart instance), it would be redundant to show the cart again.
However, if the user is on the past order page where they cannot see any representation of the cart, it would be useful to show the cart to ensure the user understands the action being performed.

\subsubsection{Runtime}
Like normal React or React-Native apps, when users interact with buttons and visual controls, the app calls the corresponding methods to update data in the state instances.
In turn, the state instances trigger the \texttt{GenieComponent}s to update their UI\@.

As shown in the bottom right of \autoref{fig:overview}, the multimodal interactions are handled through developer modules (\autoref{sec:developer-modules}), the \mrd{} modules (\autoref{sec:dsl-design}), and the generated modules (\autoref{sec:generated-modules}), collectively.
When the user touches the microphone button on the UI, \mr{} starts listening to the user's voice command and intercepts all touch events on the screen.
From this, we gather two inputs: the user's voice command and the touch point(s).
We use speech recognition from Azure to transcribe the user's speech to text.
The voice command transcript is then passed to the \textit{semantic parser} module to generate the \mrd{} code.
The touch point(s) are passed to the \textit{input UI mapping} module to determine which component and state instance the user can refer to.
Both pieces of information are then passed to the \mrd{} interpreter to execute the \mrd{} code with the corresponding relevant state instance.
\mr{} uses the methods and properties of the developer-provided \textit{state classes} to execute the \mrd{} code.
After the execution, we record both the final return value and the intermediate values during execution.
\mr{} uses the return value and the parsed DSL to generate a text response using the \textit{response generator}.
\mr{} also passes execution steps to the \textit{output UI mapping} module to determine whether and how to render the result on the screen.
Finally, the text response and the rendered UI are used to generate \textit{Feedback in Text} and \textit{Content in UI}.

\section{Framework Evaluation}

We first evaluate the development framework by checking whether our design goals have been reached.

\aptLtoX[graphic=no,type=html]{
\begin{enumerate}
\item[\textbf{F-RQ1}] How expressive is the \mr{} framework?~(\ref{f:allow-full-control}) \label{frq:expressiveness}
\item[\textbf{F-RQ2}] How much time is needed for \textit{expert developers} to develop multimodal apps using \mr{} compared with existing frameworks?~(\ref{f:maximize-reuse}) \label{frq:cost}
\item[\textbf{F-RQ3}] How easy is it to learn and use \mr{} to develop multimodal apps for \textit{novice developers}?~(\ref{f:ease-learning-curve}) \label{frq:learn}
\end{enumerate}
}{
\begin{enumerate}[label=\textbf{F-RQ\arabic*}, leftmargin=*]
\item How expressive is the \mr{} framework?~(\ref{f:allow-full-control}) \label{frq:expressiveness}
\item How much time is needed for \textit{expert developers} to develop multimodal apps using \mr{} compared with existing frameworks?~(\ref{f:maximize-reuse}) \label{frq:cost}
\item How easy is it to learn and use \mr{} to develop multimodal apps for \textit{novice developers}?~(\ref{f:ease-learning-curve}) \label{frq:learn}
\end{enumerate}
}

\subsection{Expressiveness of the Framework}

\begin{figure*}
    \centering
    \includegraphics[width=\linewidth]{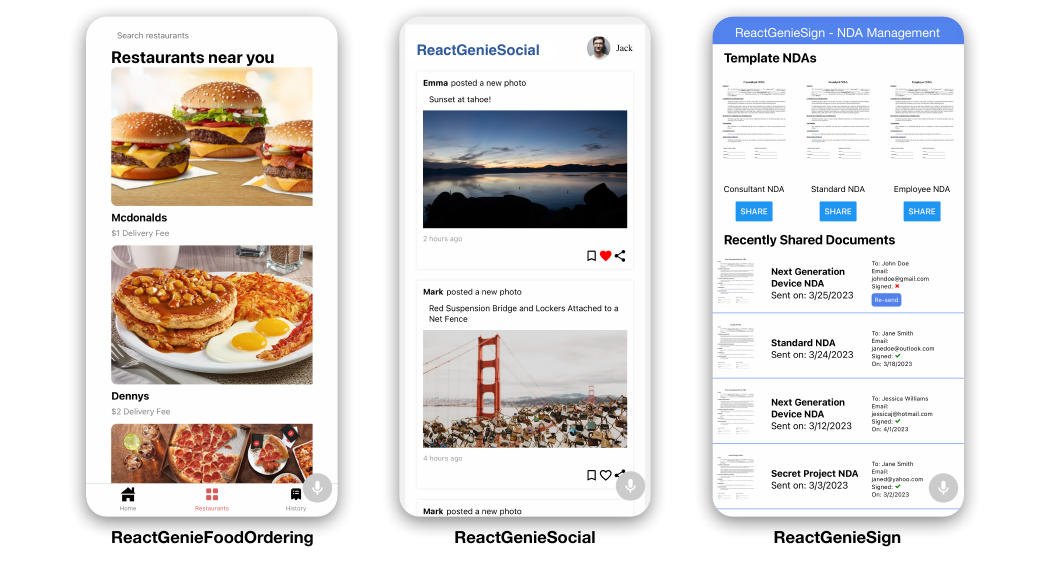}
    \caption{
    Example apps built with \mr{}.\enspace
    Left: \mr{}FoodOrdering, a food ordering app.\enspace
    Middle: \mr{}Social, a social networking app.\enspace
    Right: \mr{}Sign, a business app for distributing and collecting signed NDAs.\enspace
    }
    \Description{
        (Left) \mr{}FoodOrdering:
        It shows ``Restaurants near you'' with a list of restaurants.
        The bottom bar reads ``Home'', ``Restaurants'', and ``History''.
        (Middle) \mr{}Social:
        It shows a list of posts with pictures and text.
        Each post has a ``Bookmark'', ``Like'', and ``Share'' button.
        (Right) \mr{}Sign:
        It shows shows three NDA templates in a row.
        Each template has a ``Share'' button.
        Then, it shows a list of ``Recently shared'' NDAs.
        Each NDA lists the name, time of sharing, the recipient's email, and the NDA's status.
    }
    \label{fig:example-apps}
\end{figure*}
\begin{table*}
    \centering
    \begin{tabular}{l|lll}
        \toprule
        \textbf{App Name} & \textbf{FoodOrdering} & \textbf{Social} & \textbf{Sign} \\
        \midrule
        \texttt{DataClass} & Order, FoodItem, Restaurant & Post, User, Message & User, Document, SignatureRequest \\
        \texttt{HelperClass} & OrderItem &  & EmailAddress \\
        \texttt{GenieComponent} & 11 & 7 & 6 \\
        OtherComponents & 8 & 2 & 3 \\
        \texttt{GenieFunction} & 22 & 8 & 11 \\
        \texttt{GenieProperty} & 18 & 10 & 16 \\
        State Code (lines) & 835 & 449 & 421 \\
        Component Code (lines) & 1854 & 585 & 446 \\
        Examples (count) & 11 & 6 & 6 \\
        \bottomrule
    \end{tabular}
    \caption{
        Implementation statistics for demo apps.
        We listed all the \texttt{DataClass}, \texttt{HelperClass}, and the number of \texttt{GenieComponent} and \texttt{GenieFunction} used in the apps.
        We also listed the number of lines of code for the state and component code and the number of example parses provided for the voice parser.
    }
    \Description{
        We listed all the \texttt{DataClass}, \texttt{HelperClass}, and the number of \texttt{GenieComponent} and \texttt{GenieFunction} used in the apps.
        We also listed the number of lines of code for the state and component code and the number of example parses provided for the voice parser.
    }
    \label{tab:app-stats}
\end{table*}

To answer whether \mr{} can support the expressiveness of mobile apps (\ref{frq:expressiveness}), we built three example apps across three major categories of apps: food \& drink, social networking, and business in different interface styles, as shown in \autoref{fig:example-apps}.
The implementation statistics are shown in \autoref{tab:app-stats}.

\subsubsection{\mr{}FoodOrdering}

\mr{}FoodOrdering is a food ordering app that allows users to order food from a restaurant.
It allows users to browse menus, manage shopping carts, and check order history.
For example, users can say ``Reorder my last order'', click on a food item and say ``Add three of this to my cart'', or click on the restaurant and say ``Show me the menu of this restaurant''.
The app is comprised of 2689 lines of code, with only 88 (3\%) related to building the multimodal UI.
Note that every example parse provided by the developer takes four lines of code, and every \texttt{GenieClass}, \texttt{GenieFunction}, and \texttt{GenieProperty} annotation takes just one line of code.

\subsubsection{\mr{}Social}

\mr{}Social is a social networking app that allows users to post pictures, comment on pictures, and share pictures
with friends.
It allows users to browse, interact with, and share posts.
For example, users can say ``Show me posts from John'', ``Can you show me posts from Mark that have been liked before?'', or click on the screen and say ``Share this post with Emma''.
The app is comprised of 1034 lines of code, with only 49 (5\%) related to building the multimodal UI.

\subsubsection{\mr{}Sign}

\mr{}Sign is a business app that manages NDAs and contracts.
It allows users to create documents, share documents with clients for signing, and manage clients.
For example, users can say ``Show me the signature request from John'', click on the email address and say ``Only show me requests from this email'', or click on the email address and say ``Share the document in the most recent signature request to this email''.
The app is comprised of 867 lines of code, with only 51 (6\%) related to building the multimodal UI.

\subsubsection{Summary}

Implementing the three apps in distinct domains that support a wide range of multimodal commands demonstrates the expressiveness of \mr{}.
While building these demo apps, we noticed that different UIs are naturally decomposed into components that represent different \mr{} state instances, which made it easy to decompose the UI into \texttt{GenieComponent}s.
We were also able to lay out the graphical UI in the most appropriate way and then decompose the UI layout into individual components representing different state classes for UI mapping purposes.
We also noticed that only a small fraction (5\% on average) of the code must be written to handle multimodal interactions.
This is particularly impressive since defining multimodal interactions can be intricate and typically requires substantial code to support.

\subsection{Development Time: Expert Evaluation}
\label{sec:cost-fo-dev}

To evaluate the development time required for \mr{} apps (\ref{frq:expressiveness}), we conducted an expert evaluation comparing the time to build apps with \mr{} to the time required for building similar features with baseline tools.

\subsubsection{Study Design}

There is no readily available multimodal framework that adds multimodal capabilities to a multimodal app that has comparable capabilities to ReactGenie.
Therefore, we used GPT-3 function calling\footnotemark{} and React to implement the baseline app.
\footnotetext{\href{https://web.archive.org/web/20240229092505/https://openai.com/blog/function-calling-and-other-api-updates}{https://openai.com/blog/function-calling-and-other-api-updates}}
GPT-3 function calling is a service supported by large language models that can convert natural language into function calls.
We chose it as the baseline because a developer can use it to translate users' voice commands into function calls defined in the app, which makes it the closest off-the-shelf solution to help with the multimodal app development tasks that \mr{} supports.

For the GPT-3 function-calling condition, we used a typical React-Redux architecture where there is no object-oriented state abstraction and the UI state is stored in a monolithic data store.
For example, in the function-calling condition, the developer would implement an action on the monolithic data store called \texttt{RATE\_FOOD(food\_name: string, rating: number)} and pass the same function to the GPT-3.5-turbo endpoint as a function candidate.
When the user says ``Rate the hamburger five star'', GPT returns a function call action that it wants to call \texttt{RATE\_FOOD(hamburger, 5)}, and the developer can execute the function for the monolithic state.

In the study, we asked an expert user\footnotemark{} of \mr{} to build a multimodal timer app (\mr{}Timer, \autoref{fig:developer-apps} right).
\footnotetext{The expert developer was the first author of this paper.}
The timer app allows users to create timers and start/stop timers with voice and touch.

We asked the developer to first write the part of the app that is agnostic to \mr{} (the boilerplate code), such as the React UI code, CSS files, and basic configurations.
From there, we timed how long the expert developer finished the multimodal app with \mr{}.
From the same starting point, we also timed how long the expert developer could implement similar features with GPT-3 function calling.

\subsubsection{Results}

The implementation of both versions of the app starts with a boilerplate project that contains the framework-agnostic part of the code (337 lines of code).
The expert developer took 45 minutes to complete the \mr{} multimodal app and added 159 lines of code to complete the app.
In comparison, it took the same developer 177 minutes to implement similar functionality with GPT-3 function calling and an additional 523 lines of code.
Within that period of time, 52 minutes and 166 lines of code were spent on implementing the basic react-redux-based state management.

The additional code and time for GPT-3 function calling is due to the following:
\begin{enumerate}
    \item \textbf{The developer needs to provide the function signature to the model manually}, and when the model thinks a function call is necessary, the developer needs to call the corresponding function call.
    \item \textbf{The developer needs to build extra convenient functions for GPT-3.} Since GPT-3 can only call one function at a time, it has trouble executing actions like ``Start the exercise timer'' because that involves two steps: 1) retrieve the timer called exercise, and 2) start the retrieved timer. For GPT-3 to work, the developer has to implement a function that can start a timer by its name.
\end{enumerate}

Even with the additional lines of code, the GPT-3 function calling version lacks a few significant benefits of the \mr{} version:
\begin{enumerate}
    \item It \textbf{cannot support references by touch} due to a lack of the UI mapping module. For example, it cannot support ``Start this timer.'' while tapping on a timer.
    \item It \textbf{cannot support rich commands} unless explicitly provided by the app developer. For example, it cannot support ``Pause all timers with less than two minutes left.''
    \item It \textbf{cannot navigate to the relevant page} after executing a command due to the lack of UI mapping.
    For example, when the user says ``Start the cooking timer'' while the cooking timer is not currently on the page, the app will not show the user the cooking timer visually.
\end{enumerate}

The 3.9x time used and 3.3x lines of code to implement similar features in the baseline condition show that \mr{} drastically reduced the implementation time for developers to build multimodal apps.
In addition, the missing features in the GPT version also demonstrate the usefulness of \mr{} for expressive multimodal app development.

\subsection{Usability and Learnability: Developer Studies}
To evaluate the usability and learnability of \mr{} (\ref{frq:learn}), we conducted an IRB-approved user study asking novice developers to build multimodal applications with our framework.
Considering that developers cannot complete multimodal applications within an acceptable time frame through direct API calls, our study was focused on the framework-specific part of the implementation to measure the usability of \mr{} for building multimodal apps.
We also evaluated developers' comprehension of the framework.

\begin{figure*}
    \centering
    \includegraphics[width=\linewidth]{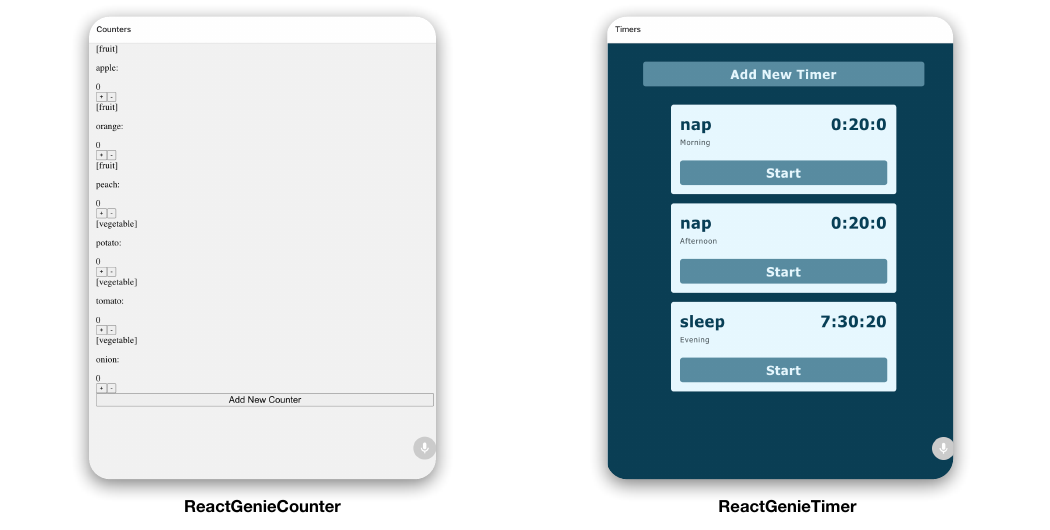}
    \caption{
        The \textbf{\mr{}Counter} and \textbf{\mr{}Timer} apps built in the developer study.
        The \textbf{\mr{}Counter} app allows users to create counters and increment/decrement counters.
        The \textbf{\mr{}Timer} app allows users to create/edit/delete timers and start/stop timers.
        Both apps support a variety of multimodal commands.
    }
    \Description{
        The \textbf{\mr{}Counter} and \textbf{\mr{}Timer} apps used in the developer study.
        The \textbf{\mr{}Counter} app allows users to create counters and increment/decrement counters.
        The \textbf{\mr{}Timer} app allows users to create timers and start/stop timers.
    }
    \label{fig:developer-apps}
\end{figure*}

\subsubsection{Study Design}

The study was facilitated using a remote desktop to ensure all participants completed the coding tasks in the same environment. 
During the study, the experimenter introduced the study goals and explained study-related concepts such as multimodal apps since most participants did not have multimodal development experience.
Then, the experimenter helped the developer connect to the remote experimental environment via video conferencing software.

The main study process contains two parts: the first, where developers learned \mr{} and the second, where they built an app in \mr{}.
In the first part, participants familiarized themselves with \mr{} by creating a multimodal counter application (\mr{}Counter, \autoref{fig:developer-apps} left) guided by a tutorial. 
In the second part, participants were asked to leverage their knowledge from the example counter app development to construct a multimodal timer application (the same \mr{}Timer app described in ~\autoref{sec:cost-fo-dev}) independently.
To make it feasible to complete the tasks within an acceptable amount of time and to ensure that participants could focus on using the \mr{} framework to implement multimodal features, we provided boilerplate code with the basic GUI implementation (similar to what was built by the expert developer in \autoref{sec:cost-fo-dev}) for the two development tasks. 

During the study, participants were required to complete a \mr{} comprehension quiz and a post-study survey that recorded their demographic information and asked them to fill out the SUS usability scale~\cite{10.1201/9781498710411-35} and the NASA-TLX cognitive load scale~\cite{10.1177/154193120605000909}.
Finally, we have a short interview about their experience using the framework.
We recorded the audio and the developer's screen during the entire process with the user's permission.
Each user received a \$60 gift card as compensation after the study.

\subsubsection{Participants}
We recruited developers with React development experience to ensure they could handle the non-multimodal-related coding tasks beyond what \mr{} is designed for.
Therefore, we designed a quiz with seven programming questions and deployed a recruitment screener.
These questions cover basic React skills (such as how to use a React component, and how to manipulate state in React) and basic object-oriented programming skills (e.g., what is \texttt{this} pointer, and how to write class constructors).
Only developers who answered at least five of the seven questions correctly were invited to participate in the study.

We recruited 12 participants (10 males and two females) in the remote user study by distributing the recruitment screener link using a convenience sample.
Our participants include student developers and professional developers with an average age of 23.8 ($\sigma = 3.43$). 
All participants had React development experience; ten developers also had TypeScript development experience.
The most experienced developer had seven years of React development experience.

\subsubsection{Results}
All participants successfully built the applications within 150 minutes.
The average completion time was 109.7 minutes ($\sigma=24.19$), with 42.3 minutes ($\sigma = 14.90$) for the first phase and 67.3 minutes ($\sigma = 17.52$) for the second phase. 
This shows that developers with React and object-oriented programming experience can quickly learn and use the \mr{} framework, illustrating the framework's ease of learning and high usability.
The user with the shortest task completion time needed only 82 minutes to complete both tasks. 

Almost all participants (11/12) answered the seven post-task quiz questions correctly. 
A single participant got one question wrong related to predefining the interface. 
This shows that developers could develop a multimodal app and correctly understand how the \mr{} code they wrote corresponds to the multimodal features, which suggests the design of the \mr{} framework is easy to understand.
The main parts that participants found relatively difficult to understand are the GUI declarations and providing example parses.
Participants said it took them more time to understand the \texttt{GenieComponent} function and how to bind data types to the corresponding interfaces.
Regarding example editing, participants thought it was hard to determine what should be included in the examples.
Although participants found these parts difficult to understand, they could use \mr{} smoothly after learning and testing their apps.
Participant 11 said, ``\mr{} incorporates natural language to UI building, and it is not hard to do that when programming.''

We evaluated \mr{}’s usability using a part of the seven-point SUS scale.
\mr{} received high SUS scores in terms of ease of use ($\mu = 6.16$, $\sigma=0.58$), enjoyable ($\mu=6.08$, $\sigma=0.79$), intuitive ($\mu=6.08$, $\sigma=0.90$), and willing to use ($\mu=6.08$, $\sigma=0.79$).
Participant 10 says, ``(\mr{} was) easy to pick up if you know React.''
At the end of the experiment, participant 12 commented that he enjoyed programming with \mr{} and asked if there was any library he could use to access \mr{} in his daily programming. 
All participants agreed that \mr{} is easy to use (medium 6 out of 7 points Likert scale), and all participants were willing to use \mr{} to build real-life multimodal applications (medium 6 out of 7 points Likert scale).
The average NASA-TLX score for the overall study was 21.99 out of 100 (the lower, the better), showing the practicality and low usage burden to program with \mr{}.

\section{Interaction Evaluation}

We evaluated the interaction provided by the \mr{} framework through two aspects: the performance of the generated neural semantic parser and the overall user experience of a \mr{} app.

\subsection{Parser Performance}

To understand how well the \mr{} parser works with information extracted from the developer's code,
we elicited commands from crowd workers for the \mr{}FoodOrdering app and tested our parser.
Specifically, we would like to know:
\aptLtoX[graphic=no,type=html]{
\begin{enumerate}
    \item RQ1: What percentage of the commands 1) are achievable with a single UI interaction on screen, 2) fall into the three targeted interactions mentioned in \autoref{sec:targeted-interactions}, or 3) are out of scope of \mr{}.
    \item RQ2: How accurate the parser is when parsing commands in the targeted interactions.
\end{enumerate}
}{
\begin{enumerate}[label=\textbf{RQ\arabic*}]
    \item What percentage of the commands 1) is achievable with a single UI interaction on screen, 2) fall into the three targeted interactions mentioned in \autoref{sec:targeted-interactions}, or 3) are out of scope of \mr{}.
    \item How accurate the parser is when parsing commands in the targeted interactions.
\end{enumerate}
}

\subsubsection{Elicitation}

We wanted to obtain multimodal commands that users may use in a real-world scenario.
We adopted a similar method as described as Cloudlicit~\cite{10.1145/3290605.3300485}.
We provided the user with three screenshots (restaurant listing page, restaurant menu page, and past orders page) of the US's two most popular food ordering apps: DoorDash and UberEats.
In our pilot study, we found that many participants' thoughts on what they can do in these apps are limited to what's on-screen and what they think the current
generation of voice assistants can do.
Therefore, we showed the final study participants 12 videos randomly, containing four videos for each of the three categories of interactions being executed on a different app (home page of the Apple app store).
Among these 12 videos, we also ensured half involved only voice and the other half contained voice and touch.

We recruited 50 participants from Prolific, a crowdsourcing platform.
We used the balanced sample options when finding participants, so we had 25 female and 25 male participants.
The age range of the participants was 20 to 79, with a median age of 29.
The survey took approximately five minutes, and we paid \$2 for each participant.

From these 50 participants, we obtained 300 commands.
We filtered out 12 unclear or unrelated responses to the survey.
For example, one participant wrote ``various good foods to order or view that can be good'' as a command.
After filtering, 288 commands remained in the dataset.

\subsubsection{\textbf{RQ1}: Percentages of categories of commands}

We classify the commands into three categories:
\begin{enumerate}
    \item \textbf{Simple UI interaction}: The command can be achieved with a single UI
    interaction on screen.
    For example, \textit{``Look at the Curry Up Now Menu''} when the restaurant is visible on screen.
    \item \textbf{Within the three targeted interaction categories}: The command falls into the three targeted interactions
    mentioned in \autoref{sec:targeted-interactions}.
    For example, one participant wrote \textit{``Order me two big macs and large fries from Mcdonald's for pickup.''}
    With a GUI, this command would typically be achieved via multiple taps to find the restaurant, add the foods, and configure the delivery options.
    \item \textbf{Out of scope of \mr{}}: The command is out of scope of \mr{}.
    For example, ``How do I repeat past orders?''.
    \mr{} tries to help people complete complex tasks, but it does not have built-in knowledge about how to use the UI of the app.
\end{enumerate}

Two researchers collaboratively labeled 30 commands to get a rubric for the rest of the commands.
They then labeled the rest of the commands (258 commands) using the rubric separately.
Both labelers labeled the same label for 224 commands and different labels for 34 commands.
Because the labels have a skewed distribution, we used Gwet's AC1~\cite{10.1348/000711006x126600} to measure the inter-rater reliability.
The AC1 score is 0.83, which means the labels are highly consistent.
They resolved the disagreement and got a final label for each command.

From this analysis, we found that 100 of the elicited commands were simple UI interactions,
172 commands fell into the three targeted interactions,
and 16 commands were out of the scope of \mr{}.

This shows that users can come up with tasks beyond just simple UI interactions even when the type of multimodal interfaces that \mr{} supports are not available in commercial apps.
It may also hint at user interest in the types of interactions that we propose here. 

\subsubsection{\textbf{RQ2}: Accuracy of the parser}
\label{sec:parser-accuracy}

We tested the parser on the 172 commands that fall into the three targeted interactions.
We ran the parser based on the \mr{}FoodOrdering app and read the generated \mrd{} to see if the parses are correct.
While working on labeling the correctness, we also noticed that many of the commands are not supported by our simple demo app, e.g., \mr{}FoodOrdering only knows delivery fees for different restaurants, but not estimated delivery times.
So we also labeled whether the feature that the command tries to use is supported by \mr{}FoodOrdering.

Our analysis showed that 101 commands are supported by \mr{}FoodOrdering, and 71 commands are not supported.
Some features that are missing from \mr{}FoodOrdering are 1) toppings/customization of a food item; 2) reviews of a restaurant or a food item; and 3) delivery time estimates for restaurants.

From the 101 commands supported by \mr{}FoodOrdering, we found that 91 commands are parsed correctly by the parser, and ten are not.
Therefore, on this dataset, the parser has an accuracy of 90\%.
A parser accuracy of 90\% should be considered very high compared to prior work on neural semantic parsers' accuracy on compound commands~\cite{10.1145/3314221.3314594}.
Seven of the ten incorrect parses would result in syntax errors (such as used \texttt{.first()} rather than \mrd{} supported \texttt{[0]}) or runtime errors (such as ordering the last meal from ``A'' has been translated to order the food called ``A'', thus the system will not be able to find ``A'' as a food).
Three of the incorrect parses would be helpful but not give the desired behavior.
For example, ``Find me the closest pizza restaurant'' was translated to ``find the closest restaurant''.
None of the errors resulted in behavior completely different from the user's expectations.

We also looked at the 71 commands that are not supported by \mr{}FoodOrdering.
These commands mention features that are not in the \mr{}FoodOrdering app.
To our surprise, the parser also generated sensible \mrd{} for the majority (38) of these commands.
The \mr{} parser approximates the request command with available features in the app for 24 of these commands.
For example, \mr{} parser generates \texttt{Restaurant.GetRestaurant(name: "pizza hut").getFoodItems().between(field: .price, from: 0, to: 5)} for the command \textit{``What deals does pizza hut have?''}.
In this case, the parser approximates deals with food items that are less than 5 USD\@.
For 14 commands, the parser would generate function calls and property accesses that are not supported by the app.
For example, the parser parsers \textit{``What time does Chipotle open?''} to \texttt{Restaurant.GetRestaurant(name: "Chipotle").openingTime}.
In this case, \mr{}FoodOrdering does not have the property \texttt{openingTime} for restaurants, but the parser is still capable enough to generate a sensible \mrd{}.
In the future, the \mr{} runtime can leverage this information to inform the user of the missing property and potentially even suggest the developer add common missing features to the app.

There were 33 unsupported commands that were not parsed correctly by the parser.
Some of them are due to the parser generating ungrammatical \mrd{}, and others use incorrect properties and methods.
For example, the parser parses \textit{``Find restaurants that deliver in less than 25 minutes.''} to \texttt{Restaurant.All().matching(field: .deliveryFee, value: < 25)}.
In this case, \mr{}FoodOrdering does not know the estimated delivery time of restaurants, but the correct parsing should be \texttt{Restaurant.All().between(field: .deliveryTime, from: 0, to: 25)}.

The results show that \mr{} parser is a reasonably good implementation for parsing natural language commands to \mrd{} using only information extracted from the shared logic code and the few-shot examples provided by the developer.

Another interesting metric is that 104 of the 172 commands contain at least one touch point, but there are only 18 cases where these touch points are required to execute the command.
In many of these commands, the user taps relevant objects, hoping that it would help the system understand.
For example, they would tap on the ``Restaurant'' menu bar while saying ``Show me a pizza restaurant near me.''
Another interesting observation is that when they referred to objects on screen, they often would not use a reference term like ``this'' or ``that.''
Instead of saying \textit{``Reorder this order''}, the participant would say \textit{``Reorder my Mendocino Farms order from Thursday.''}
This shows a potential opportunity to improve the semantic parser by always adding the touch context even when it seems unnecessary.


\subsection{Usability of Supported Interactions}

We conducted a usability study with the \mr{}FoodOrdering app to understand if the generated multimodal UIs are useful for end users.
We measured the performance of the multimodal UIs in terms of the time it takes to complete a task, the cognitive load, and the usability of the experience when using the app compared to the same app limited to using only the GUI.

\subsubsection{Study Design}

In the study, we asked participants to complete a set of tasks using two variants of the \mr{}FoodOrdering app, one
generated by \mr{} and one limited to only the GUI.
We used a within-subjects design, where each participant completed the same tasks using both variants of the app.
For each app variant, we first teach the participant how to use the app using one training task.
Specifically, for the \mr{} condition of the app, we explained that in addition to typical touch-only interaction, they can also tap the microphone button to initiate a speech + gesture command when they want to.
We then asked them to complete two test tasks with the variant.
After completing the two tasks, we asked them to complete a survey about their cognitive load using the system (using NASA-TLX~\cite{10.1177/154193120605000909}) and the usability of the experience (using SUS~\cite{10.1201/9781498710411-35}).
At the end of the study, we asked the participants about their subjective preferences between the two variants of the app and their reasons for their preferences.

We designed one training task and two test tasks for each variant of the app.
The training tasks are to order the cheapest food item from the menu of two different restaurants.
The test tasks are re-ordering an order from two different days (today or yesterday), and finding the most recent order containing two different items.
When presenting these tasks, we described a scenario, what we wanted them to do, and the expected outcome (order placed screen or a certain screen showing a past history order).
We counterbalanced the order of the two apps and the order of the three tasks.

\subsubsection{Participants}

We recruited 16 participants, aged 18--30, with a median age of 23.
Eight of our participants are female, six are male, one stated other, and one prefers not to say.
One of our participants uses food ordering daily, two use it weekly, five use it monthly, seven use it a few times per year, and one rarely or never uses it.
All of our participants use graphical mobile interfaces daily.
Two of our participants use voice interfaces daily, four use them weekly, two use them monthly, four use them a few times per year, and four rarely or never use them.
The study took about 30 minutes to complete, and we compensated each participant with a 15 USD Amazon gift card for their time.

\subsubsection{Results}

\begin{figure}
    \centering
    \includegraphics[width=3.5in]{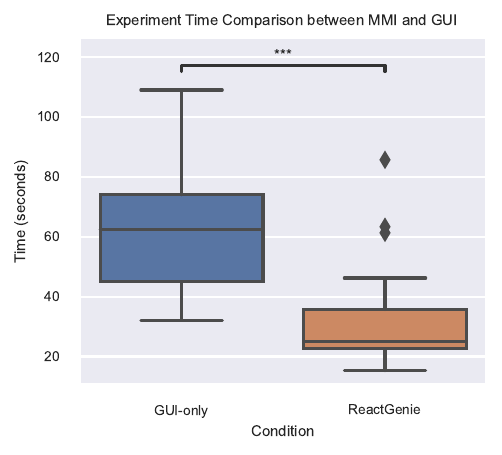}
    \caption{Users can complete common tasks faster ($p=0.0004, t=3.955$) with the multimodal interface (MMI) app built with \mr{} compared with a baseline GUI app.}
    \Description{
        The average time it takes to complete both tasks using the graphical UI is 63.6 seconds, while the average time it takes to complete a task using the multimodal UI is 33.6 seconds.
        We used a paired t-test and found that the difference is statistically significant ($p=0.0004, t=3.955$).
    }
    \label{fig:time}
\end{figure}
\begin{figure*}
    \centering
    \includegraphics[width=\linewidth]{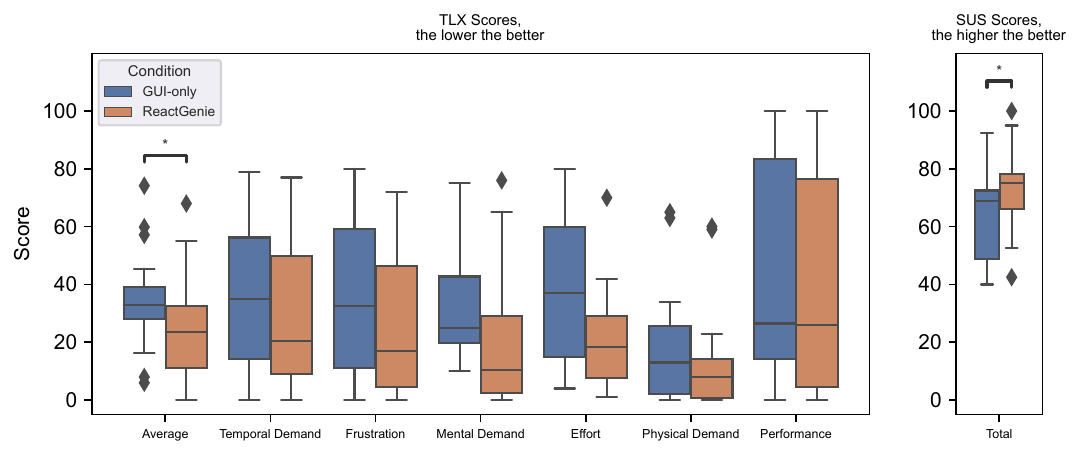}
    \caption{Cognitive load (left) and usability (right) of the GUI vs. the \mr{} multimodal UI.}
    \Description{
        The left figure shows NASA-TLX scores for the GUI and the multimodal UI.
        It shows that the average cognitive load for the multimodal UI is lower than the GUI with statistical significance.
        The right figure shows the usability scores for the GUI and the multimodal UI.
        It shows that the average usability score for the multimodal UI is higher than the GUI with statistical significance.
    }
    \label{fig:survey}
\end{figure*}

We computed the time it took to complete each task using the graphical UI and the multimodal UI (see \autoref{fig:time}).
The average time it took to complete each task using the graphical UI was 63.6 seconds, while the average time it took to complete each task using the multimodal UI was 33.6 seconds.
We used a paired t-test and found that the difference is statistically significant ($p=0.0004, t=3.955$).

We compared NASA-TLX average scores between the two conditions (see \autoref{fig:survey} left).
The average NASA-TLX score for the graphical UI is 34.5, while the average NASA-TLX score for the multimodal UI is 24.6 (note: lower is better).
We used a Wilcoxon test and found that the difference is statistically significant ($p=0.013, z=21$).

We compared the average SUS scores between the two conditions (see \autoref{fig:survey} right).
The average SUS score for the graphical UI is 63.3, while the average SUS score for the multimodal UI is 73.0 (note: higher is better).
We used a Wilcoxon test and found that the difference is statistically significant ($p=0.031, z=22$).

\textit{11 out of 16 of our participants preferred the \mr{}} generated multimodal UI over the graphical UI\@.
For participants who preferred the multimodal UI, the most common reason was that it was easier to use (P4, P8, P13, P16).
P2 mentioned that they would prefer to use a mix of both in the real world, which is well supported by \mr{}.
P6 mentioned that the multimodal UI could be especially useful when they are unfamiliar with the app.
P12 mentioned that the multimodal commands allowed them to do more complex tasks with a clear path rather than searching and finding out how to do so in the graphical UI\@.
For participants who preferred the graphical UI, the most common reason was that the speech recognition was not accurate (P5, P7, P14).
P9 and P11 mentioned that they generally do not use voice interfaces.

\subsubsection{Discussion}

The results of our usability study show that the multimodal UIs generated by \mr{} are more efficient, have a lower cognitive load, and have higher usability compared to the corresponding graphical UI versions.
These findings suggest that the \mr{} system is successful in generating multimodal UIs that enhance the user experience, making it easier and more efficient for users to complete tasks.
All participants, when using the \mr{} variant of the app, used both typical touch interaction for simple navigation and browsing and touch (optional) + speech interaction for more complex inputs.
The combination of graphical and voice interfaces allows users to take advantage of the strengths of each modality, resulting in a more streamlined and enjoyable experience.

\section{Discussion}

In this section, we will characterize the properties of the multimodal interaction supported by \mr{}, and discuss the limitations, future work, safety, and implications of \mr{}.

\subsection{Properties of Multimodalities in \mr{}}

We can characterize the properties of the \mr{} interactions using the framework proposed by \citet{10.1007/978-1-5041-2896-4_19}.
\mr{}'s voice + (optional) touch actions are implemented using the same functions used for graphical user interfaces.
Therefore, all voice and touch actions are equivalent, and almost no actions belong to the assignment category, meaning that almost no actions can only be performed using a specific modality.
Note that the functions only interact with the state.
Commonly minute actions such as scrolling will not be included in part of the state, so actions like ``Scroll to here'' will not be supported by the voice + touch commands.
This is intentional, as scrolling is likely more efficient using touch than voice + touch.
However, in \mr{}, developers can expose anything as part of the state.
In an e-book or a map application, where the current read position/map position is a crucial part of the experience, developers can choose to expose the position as part of the API.
The user's command ``scroll to here'' can be translated to \texttt{ReadingPosition.SetPosition(position: ReadingPosition.Current())}.

For redundancy, \mr{} apps will only accept touch commands when the microphone button is not activated and will only accept voice + touch commands when the microphone button is activated to achieve partial redundancy.
For complementarity, \mr{} primarily supports voice for actions and touch for deictic gestures for reference.
After the user clicks the microphone button, the order of touch and voice does not matter, the user can tap first then speak, speak first then tap, or tap while speaking.
The voice and touch mode will end automatically after receiving no new words or touches for 0.5 seconds.

\subsection{Limitations and Future Work}

\mr{} is the first attempt at integrating multimodal development into the modern declarative GUI development process.
It provides a familiar workflow, allows the reuse of state code and UI, and can understand rich multimodal commands.
However, it is far from perfect.
There are three directions that future work can improve on 1) better voice interfaces, 2) better developer support, and 3) support for more modalities.

\subsubsection{Better Voice Interfaces}

\mr{} accepts the user's voice and touch input and generates text and GUI output based on the result.
We currently provide text but not voice feedback, which is easy to change by using a commercial text-to-speech module.
However, a more significant area of improvement is in maintaining natural language context.
For example, if the user says, ``What is the best pizza restaurant?'' and then asks, ``What about Chinese food?'', the system should be able to understand that the user is asking about the \textit{best} Chinese food instead of any Chinese food restaurant.
Note that \mr{} can handle some conversations gracefully by using the current UI context as the context for the next command.
An example would be the user saying, ``Find me the cheapest hamburger at McDonald's'' and then asking, ``Order one of that'' (\mrd{}: \texttt{Order.GetActiveCart().addItems([FoodItem.Current()])}).
\mr{} would present the food item after the first command, and when the user says the second command, \mr{} would know that the user is referring to the food item presented in the GUI\@.

Another way to improve the reliability of the generated interfaces is to better leverage multimodal commands for disambiguation.
As shown in \autoref{sec:parser-accuracy}, many of our elicited commands include redundant information from voice and the GUI\@.
Future work can leverage this redundancy and provide extra GUI context to the semantic parser to further push the parser's accuracy closer to 100\%.
For example, a potential path is to improve \mr{} runtime to extra GUI context and enhance the neural semantic parser through prompt engineering to allow it to process the extra GUI context.

\subsubsection{Better Developer Support}

Although \mr{} provides a customizable and easy way of programming multimodal apps, it can still be improved.
One area that we see as a potential improvement is to reduce the number of examples necessary and to increase the effectiveness of the example parses.
The majority of these examples are there for teaching the parser how to generate syntactically correct \mrd{} code.
However, given we have the interpreter, we can potentially use it as an example generator to teach the parser how to generate syntactically correct \mrd{} code, similar to the method used in SEMPRE~\cite{sempre} or Genie~\cite{10.1145/3314221.3314594}.
Another route is to fine-tune the Codex model with the \mrd{} code generated by the interpreter so that the interpreter can generate syntactically correct \mrd{} code with fewer examples.

Future extensions to the \mr{} framework can also help developers identify potential voice commands that the user may want to say.
Using \mr{}FoodOrdering as an example, its API only supports 59\% of the commands that we elicited from crowd workers.
Some top categories of unimplemented commands are about delivery time (mentioned in 8 commands), food customization options (8), discount/deal information (7), pickup/delivery support of restaurants (6), food types (e.g.,\ vegetarian or vegan) (6), and calorie/health/allergy information (6).
If we can implement these commands, we can potentially reduce the number of unsupported commands by more than 50\%.
Future work can consider embedding elicitation studies directly into the app development cycle, or the framework could record unsupported commands from actual users and use this data as feedback to the development team to help improve the system after the initial deployment.

\subsubsection{Support More Modalities}

As stated in \autoref{sec:targeted-interactions}, \mr{} is targeted at gesture and voice interactions, and it is optimized for deictic gestures used for reference.
Currently, \mr{} can support complex gestures in the regular GUI provided other gesture interaction frameworks, but not simultaneous voice + complex gesture interactions.
Future frameworks can improve on supporting more diverse gestures and modalities beyond these two categories.

Gestures and modalities other than voice can also be used to indicate the actions that the user wants to perform.
This can be supported by summarizing what happened using other gestures in language and presenting them along with the user's speech commands.
For example, suppose a user says, ``Animate this box,'' and performs a move while rotating and gesturing on the screen.
In that case, we can present \texttt{voice: animate this box} and \texttt{gesture: rotate and move from x1, y1, to x2, y2} to the large language model and ask the model to generate \mrd{} for this animation.
(\texttt{x1, y1} is the coordinate of the gesture start coordinate, and the \texttt{x2, y2} is the gesture end coordinate.)

Another way gestures can be integrated is through a new class called \texttt{Gesture} provided to the \mr{} system to indicate the gesture trajectory that the user performed.
For example, if the user says ``Deploy soldiers along this path'' in a QuickSet-like~\cite{10.1145/266180.266328} system, \mr{} can translate it to \texttt{Solders.DeployOnPath(path: Path.FromGesture(gesture: Gesture.Current()))}.

Through a combination of the aforementioned methods and utilization of the gesture recognition/modality understanding components of prior multimodal interaction libraries~\cite{10.1145/2070481.2070500,10.1145/1358628.1358881}, future work can combine the flexibility and expressiveness of \mr{} with different modalities.

\subsection{Takeaways for Future UI Frameworks}

\mr{} demonstrated a feature-first, rather than interaction-first, way of implementing multimodal apps.
\mr{} lets developers do what they know best: implement the features the users want, and the framework intelligently takes care of users' diverse interactions using their specific way of making commands.
This is similar to modern graphical UI frameworks that, rather than directly passing the user's touch point and letting developers implement the rest, ask developers to provide functions and pages at a feature level.
The graphical UI framework handles how to render and convert user touch points and typical gestures to the corresponding function calls.
This was not easy to achieve previously for multimodal interactions.
\mr{} made it possible because of the power of LLM's language-parsing capability and the expressiveness of \mrd{}.
Future frameworks can build on this idea to support more modalities and be more adaptive to the user's interaction context.

Other researchers in intelligent human interfaces can also take advantage of how \mr{} integrates LLMs into the human interaction loop.
Rather than asking developers to call LLMs in their app, which requires them to understand prompting techniques, \mr{} automatically generates an interface to the LLM as a programming framework.
This program-LLM interface provides the available functionality to the LLM and allows the LLM to call the compound functionality of the program.
Other intelligent human interfaces, such as conversational agents or adaptive interfaces, can also learn from this technique.
They can automatically generate an interface from the developer's code that exposes available features and relevant context to LLMs and allows LLMs to use a programming language (maybe \mrd{}) to perform their job.

\subsection{Safety and Implications}

\mr{} uses a machine learning model to understand the users' commands.
This may introduce safety issues when the wrong command is interpreted and executed.
In our evaluation, a pleasant finding is that wrongly parsed commands are either not executable or still helpful towards reaching the user's intended goal.
Further risks can be reduced by having a more accurate semantic parser.

Also, compared with an end-to-end natural language assistant like ChatGPT\footnote{\href{https://web.archive.org/web/20240228110251/https://openai.com/blog/chatgpt/}{https://openai.com/blog/chatgpt/}}, \mr{} allows more control over the \textit{presented information} and \textit{performed actions}.
\mr{}'s framework only processes and shows information in the developer's provided state code and can reduce hallucinated information.
One particular case of error in our testing was when the user asked for the delivery time, but because the app does not support delivery time estimation, \mr{} returned the delivery fee instead.
In this case, the text feedback mechanism can be used to inform the user of the information that is returned.
In the future, an error correction mechanism would be useful for the user to report the error, and this may allow the developer to fix it.

For performed actions, \mr{} gives text feedback and renders the related UI elements to ensure the user is aware of the command being executed, so when there is an error, the user can easily identify and recover from it.
A design decision we made while creating the three demo apps is not to expose non-recoverable actions to voice.
For example, in the \mr{}FoodOrdering app, the user can browse items, add items to the cart, and go to the checkout page via voice, but placing the order will only present the checkout page and require the user to click the ``Place Order'' button to place the order.
This way, the irreversible action is only triggered through the GUI, with little room for error.
It should be strongly recommended to developers using \mr{} to either not expose (via Genie annotations) functions that would create an irreversible effect, such as payment-related or account management functions, or add a confirmation stage via graphical user interfaces or explicit voice commands.

Another implication of \mr{} is the possible negative social implications of noisy multimodal interaction.
\mr{} encourages users to use voice and touch to quickly achieve their goals without going through multiple UI actions and exploration steps.
The benefit of \mr{} comes from the expressiveness of voice and touch, but voice interfaces may not always be appropriate.
One possibility is to explore silent voice interfaces like those presented by \citet{Denby2010} that can be used in public spaces.

\section{Conclusions}

Commercial user interfaces have stagnated with the GUI for more than a decade.
Although these GUIs work well for communicating exact information (e.g., from a menu) and binary actions (e.g., using a button), they are not expressive enough to communicate and collect rich multimodal information, such as the way a waiter or waitress can obtain a person's order from a restaurant menu.
\mr{} attempts to break that UI stagnation by enabling developers to create multimodal UIs that allow for more expressiveness than traditional GUIs, with little additional programming effort.
\mr{} accomplishes this first by introducing an interaction programming paradigm where the interaction logic is better separated from user interface implementation and second by using a powerful natural language understanding module that leverages the capabilities of LLMs to execute code in the interaction logic.
In this paper, we demonstrated the easy adoption of the \mr{} framework for developers and tested the expressiveness, usefulness, and accuracy of the resulting multimodal apps with end-users.
In the future, by introducing developer tools based on frameworks like \mr{}, and the research on the multimodal interactions these tools enable, we hope to see humans communicating with computers more expressively and more easily.

\begin{acks}
    We would like to thank the reviewers for their insightful feedback and the participants of our user studies for their invaluable input.
    We also want to acknowledge Meta Platforms, Inc., the Alfred P. Sloan Foundation, and the Verdant Foundation for their generous financial support.
    We are also grateful to Microsoft for providing Azure AI credits, which have been instrumental in advancing our research.
\end{acks}

\bibliographystyle{ACM-Reference-Format}
\bibliography{main-fixed, extra}

\appendix
\section{Grammar of \mrd{}}
\label{sec:grammar}
\begin{align*}
\textit{top} \Coloneqq & \textit{value} \mid \textit{all\_symbol} \\
\textit{all\_symbol} \Coloneqq & \textit{index\_symbol} (\text{.} \textit{all\_symbol})? \\
\textit{index\_symbol} \Coloneqq & \textit{function\_call} \mid \textit{symbol} (\text{[} \textit{int\_literal} \text{]})? \\
\textit{function\_call} \Coloneqq & \textit{symbol} (\text{(} \textit{parameter\_list}? \text{)}) \\
\textit{parameter\_list} \Coloneqq & \textit{parameter\_pair} (\text{,} \textit{parameter\_pair})* \\
\textit{parameter\_pair} \Coloneqq & \textit{symbol} \text{:} \textit{value} \\
\textit{value} \Coloneqq & \text{true} \mid \text{false} \mid \textit{int\_literal} \mid \textit{float\_literal} \mid \textit{all\_symbol} \mid \\ & \textit{accessor} \mid \text{"} \textit{string} \text{"} \mid \text{[} \textit{array\_value} \text{]} \\
\textit{accessor} \Coloneqq & \text{.} \textit{value} \\
\textit{array\_value} \Coloneqq & \textit{value} (\text{,} \textit{value})* \\
\textit{symbol} \Coloneqq & \textrm{[}a-zA-Z\textrm{\_][}a-zA-Z0-9\_\text{]} *  \\
\textit{int\_literal} \Coloneqq & (\text{+} \mid \text{-})? [0-9]+ \\
\textit{float\_literal} \Coloneqq & (\text{+} \mid \text{-})? [0-9]*\text{.}[0-9]+
\end{align*}

\section{Example parses}

Here is a code excerpt from the developer-provided example parses for ReactGenieFoodOrdering:

\begin{lstlisting}
Order.Examples = [
    {
        user_utterance: "What is the total price of my order?",
        example_parsed: "Order.GetActiveCart().getTotalPrice()",
    },
    {
        user_utterance: "Order a burger and two fries.",
        example_parsed: "Order.GetActiveCart().addItems(items: [OrderItem.CreateOrderItem(foodItem: FoodItem.GetFoodItem(name: \"burger\")), OrderItem.CreateOrderItem(foodItem: FoodItem.GetFoodItem(name: \"fries\"), quantity: 2)])",
    },
    {
        user_utterance: "I would like to place an order for pick up",
        example_parsed: "Order.GetActiveCart().setPickUp(pickup: true)",
    },
    {
        user_utterance: "What's the cheapest item in my order?",
        example_parsed: "Order.GetActiveCart().items.sort(field: .foodItem.price(), ascending: true)[0]",
    },
    {
        user_utterance: "What have I ordered last time from mcDonalds?",
        example_parsed: "Order.OrderHistory().matching(field: .restaurant, value: Restaurant.GetRestaurant(name: \"mcDonalds\"))[0].items"
    }
]
\end{lstlisting}

\section{Prompt for \mrd{} parser}
Here is an example prompt for OpenAI Codex to convert the user command to \mrd{} and its expected response.
Text snippets starting with \texttt{//} are sent to the LLM as part of the prompt.
It resembles code comments to help the LLM understand the structure of the prompt.
To help the reader understand where different data are generated, we added comments surrounded by \texttt{<>}, which are not part of the prompt sent to the LLM.

\begin{lstlisting}[language=Prompt]
// Here are all the functions that we have
<developer's class skeletons>
class Restaurant {
    string name;
    string address;
    string cuisine;
    float rating;
    
    // All active restaurants
    static Restaurant[] All();
    
    // The current restaurants
    static Restaurant Current();
    
    // Get a list of foods representing the menu from a restaurant
    Food[] menu;
    
    // Book reservations on date
    Reservation get_reservation(date: DateTime)
}
...
// Examples:
<example parses>
user: get me the best restaurant in Palo Alto
parsed: Restaurant.all().matching(field: .address, value: "Palo Alto").sort(field: .rating, ascending: false)
...

// Current User Interaction
<current user command>
user: order the same burger that I ordered at McDonald's last time
parsed:
<expected LLM response>
Order.Current().addFoods(foods: Order.All().matching(field: .restaurant, value: Restaurant.All().matching(field: .name, value: "McDonald's")).sort(field: .orderTime, ascending: false)[0].foods)
\end{lstlisting}

\end{document}